\documentclass[%
 reprint,
 amsmath,amssymb,
 aps,
prb,
 superscriptaddress
]{revtex4-1}

\usepackage{graphicx}
\usepackage{dcolumn}
\usepackage{bm}
\usepackage{braket}
\usepackage{color}

\begin{document}

\preprint{APS/123-QED}

\title{Topological semimetal phase with exceptional points in one-dimensional non-Hermitian systems}

\author{Kazuki Yokomizo}
\affiliation{Department of Physics, Tokyo Institute of Technology, 2-12-1 Ookayama, Meguro-ku, Tokyo, 152-8551, Japan}
\author{Shuichi Murakami}
\affiliation{Department of Physics, Tokyo Institute of Technology, 2-12-1 Ookayama, Meguro-ku, Tokyo, 152-8551, Japan}
\affiliation{TIES, Tokyo Institute of Technology, 2-12-1 Ookayama, Meguro-ku, Tokyo, 152-8551, Japan}%




%
\begin{abstract}
Energy bands of non-Hermitian crystalline systems are described in terms of the generalized Brillouin zone (GBZ) having unique features which are absent in Hermitian systems. In this paper, we show that in one-dimensional non-Hermitian systems with both sublattice symmetry and time-reversal symmetry such as the non-Hermitian Su-Schrieffer-Heeger model, a topological semimetal phase with exceptional points is stabilized by the unique features of the GBZ. Namely, under a change of a system parameter, the GBZ is deformed so that the system remains gapless. It is also shown that each energy band is divided into three regions, depending on the symmetry of the eigenstates, and the regions are separated by the cusps and the exceptional points in the GBZ.
\end{abstract}
\pacs{Valid PACS appear here}
\maketitle
%
%

\section{Introduction\label{sec1}}
Non-Hermitian quantum mechanics has been attracting much attention in many fields of physics in the past decades. Many experimental studies have realized various physical systems with non-Hermitian effects~\cite{Eichelkraut2013,Xu2016,Xiao2017,Bahari2017,Bandres2018,Rosenthal2018,Li2019,Kremer2019,Wu2019,Brandenbourger2019,Sakhdari2019,Tuniz2019,LXiao2019,Xiao2020,Poli2015,Zeuner2015,Weimann2017,St2017,Parto2018,Pan2018,Ghatak2019e,Helbig2020}. Among these experimental studies, appearance of exceptional points and rings where some energy eigenvalues become degenerate and the corresponding eigenstates coalesce~\cite{Berry2004,Heiss2012} and intriguing phenomena have been observed~\cite{Dembowski2001,Zhen2015,Ding2016,Zhou2018,Ding2018,Cerjan2019,XZhang2019,Guo2009,Feng2013,Brandstetter2014,Chen2017,XWang2019}. At such non-Hermitian degeneracy, since the Hamiltonian is nondiagonalizable, these degeneracies are unique to non-Hermitian systems. As is motivated by these experimental studies, the existence of exceptional points, rings, and surfaces and phenomena induced by them have been theoretically predicted in various physical systems~\cite{Ding2015,Lee2016,Lin2016,Leykam2017,Xu2017,Gonzalez2017,Zyuzin2018,Wang2018,Martinez2018,Molina2018,Jin2018,Carlstrom2018,Pan2019,Moors2019,Budich2019,Yang2019,Zhou2019,Zhu2019,Hatano2019,Chen2019,Zhang2019,Zhong2019,Carlstrom2019,Zyuzin2019,Xiao2019,Pan2019v2,Rui2019,Kimura2019}, and under some symmetries they are classified in terms of topology~\cite{Okugawa2019,Kawabata2019v2,Yoshida2019,Lin2019,Yoshida2019v2}.

Recent theoretical studies have been focusing on topological systems in solid state physics~\cite{Hu2011,Esaki2011,Kozii2017,Harari2018,Shen2018,Yoshida2018,Gong2018,Philip2018,Chen2018,Liu2019,Papaj2019,Ghatak2019t,McClarty2019}. The bulk-edge correspondence has been notably under debate~\cite{Ye2018,Malzard2018,Yao2018v2,Kawabata2018v2,Takata2018,Kawabata2019,Bliokh2019,Wang2019,tLiu2019,Edvardsson2019,KLZhang2019,Lieu2019,Yokomizo2019,Okuma2019,HWu2019,Brzezicki2019,Kawabata2019v3,Borgnia2020,Okuma2020,Yin2018,Herviou2019,YWu2019,RChen2019,LLi2019,Loic2019,Wang2020,Runder2009,Liang2013,Zhu2014,Zhao2015,Jin2017,Yuce2018pra,Lieu2018,Klett2018,Yuce2018,Ge2019,Fu2020,Kunst2018,Yao2018,Jin2019,Lee2019,Kunst2019,Deng2019,Imura2019,Song2019,Song2019v2,Longhi2019,KZhang2020} since it seems to be violated in contrast to Hermitian systems. The main issue of the bulk-edge correspondence in non-Hermitian systems is that there is a difference between the energy eigenvalue in a periodic chain and that in an open chain. This difference is caused by the non-Hermitian skin effect~\cite{Yao2018}. In Refs.~\onlinecite{Yokomizo2019} and \onlinecite{Yao2018}, it was shown that while the Bloch wave number $k$ takes a real value in a periodic chain, it becomes complex in an open chain, and that the value of $\beta\equiv{\rm e}^{ik}$ is confined on a loop on the complex plane so that continuum bands are reproduced in a large open chain. Then the loop of $\beta$ is called the generalized Brillouin zone (GBZ), denoted as $C_\beta$, which is a generalization of the Brillouin zone in Hermitian systems. We note that $C_\beta$ is deformed as system parameters change and that it can have cusps~\cite{Yokomizo2019}. As a result, one can establish the bulk-edge correspondence between a topological invariant defined in terms of $C_\beta$ and existence of the topological edge states.

In this paper, we show that in one-dimensional (1D) non-Hermitian systems with both sublattice symmetry (SLS) and time-reversal symmetry (TRS), a topological semimetal (TSM) phase becomes stable under a continuous change of system parameters. This TSM phase appears as an intermediate phase between a normal insulator (NI) phase and a topological insulator (TI) phase characterized by a topological invariant. Furthermore, we find that each continuum band is divided into three regions, depending on the symmetry of the eigenstates, and that these regions correspond to energy eigenvalues being real, pure imaginary, and general complex values. These two features are closely related, and they are unique features due to non-Hermiticity. As system parameters change continuously, $C_\beta$ is deformed so that gapless points always lie on $C_\beta$, and the system remains gapless. Moreover, $C_\beta$ is divided into the three regions, whose boundaries are given by cusps on $C_\beta$ and exceptional points.

%
%

\section{Non-Hermitian Su-Schrieffer-Heeger model\label{sec2}}
\begin{figure}[]
\includegraphics[width=8.5cm]{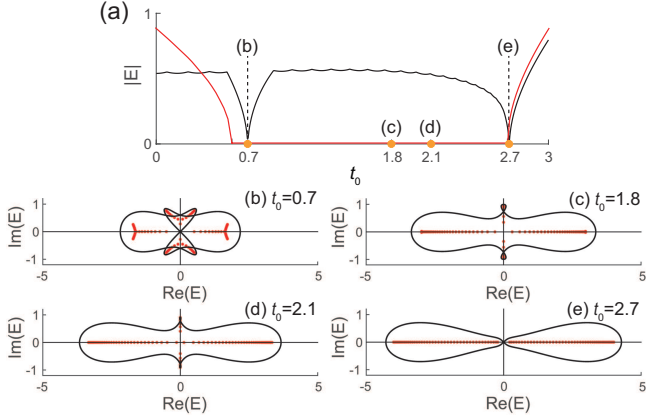}
\caption{\label{fig1} (a) Energy gap in the non-Hermitian Su-Schrieffer-Heeger model with a periodic boundary condition (black) and with an open boundary condition (red). Because of the sublattice symmetry, the two energy eigenvalues $E_1$ and $E_2$ are opposite in sign, $E_1=-E_2$, and the plot shows their absolute values $\left|E_i\right|~(i=1,2)$. (b)-(e) Energy eigenvalues in a periodic chain (black) and in an open chain (red), respectively. The values of the parameters are $t_1=1.2,t_{-1}=0.5,\Delta_1=0.3$, and $\Delta_{-1}=-0.7$.}
\end{figure}
To show the above features, we introduce the non-Hermitian Su-Schrieffer-Heeger (SSH) model. Although this model has been much studied in many previous works~\cite{Poli2015,Zeuner2015,Weimann2017,St2017,Parto2018,Pan2018,Ghatak2019e,Helbig2020,Yin2018,Herviou2019,YWu2019,RChen2019,LLi2019,Loic2019,Wang2020,Runder2009,Liang2013,Zhu2014,Zhao2015,Jin2017,Yuce2018pra,Lieu2018,Klett2018,Yuce2018,Ge2019,Fu2020,Kunst2018,Yao2018,Jin2019,Lee2019,Kunst2019,Deng2019,Imura2019,Song2019,Song2019v2,Longhi2019}, most works focused only on special cases, while some general and important features were left unexplored. We comment on this point in detail in Appendix~\ref{secB}. Now the real-space Hamiltonian of this system is written as
\begin{eqnarray}
H&=&\sum_n\left(t_1^+c_{n,{\rm A}}^\dag c_{n+1,{\rm B}}+t_0^+c^\dag_{n,{\rm A}}c_{n,{\rm B}}+t_{-1}^+c_{n+1,{\rm A}}^\dag c_{n,{\rm B}}\right. \nonumber\\
&+&\left.t_1^-c_{n,{\rm B}}^\dag c_{n+1,{\rm A}}+t_0^-c^\dag_{n,{\rm B}}c_{n,{\rm A}}+t_{-1}^-c_{n+1,{\rm B}}^\dag c_{n,{\rm A}}\right),
\label{eq1}
\end{eqnarray}
where all the parameters are set to be real. Then the system preserves the SLS and TRS, which are defined as $\Gamma H\Gamma^{-1}=-H,~{\cal T}H^\ast {\cal T}^{-1}=H$, where $\Gamma$ and ${\cal T}$ are unitary matrices, and $\Gamma^2={\cal T}{\cal T}^\ast=+1$. Henceforth we set all the parameters as $t_0^\pm=\mp t_0$, $t_1^\pm=t_1\pm\Delta_1$, and $t_{-1}^\pm=t_{-1}\mp\Delta_{-1}$. Here, in this system, we show the energy eigenvalues in a periodic chain and in an open chain in Figs.~\ref{fig1}(b)-\ref{fig1}(e), showing a qualitative difference between these two chains. In particular, in a long open chain, the gapless phase extends over a certain region in the parameter $t_0$ as shown in Fig.~\ref{fig1}(a). While the qualitative difference of the energy eigenvalues between a periodic chain and an open chain has been discussed in various non-Hermitian systems, this appearance of the stable gapless phase only in an open chain is unexpected in 1D systems.

\begin{figure}[]
\includegraphics[width=8.5cm]{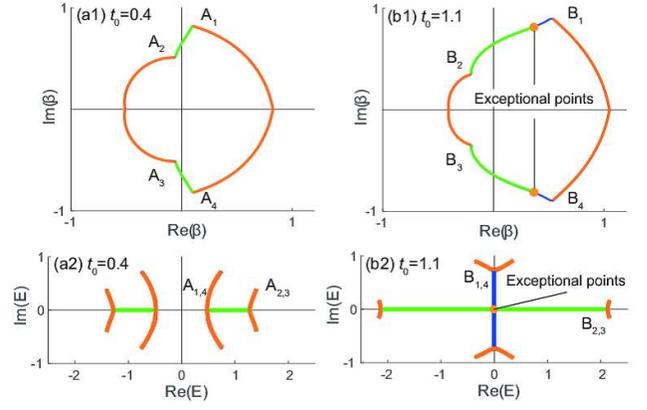}
\caption{\label{fig2}Generalized Brillouin zone (GBZ) and continuum bands in the non-Hermitian Su-Schrieffer-Heeger model. TRS-unbroken region, STS-unbroken region, and TRS/STS-broken region are shown in green, in blue, and in orange, respectively. The values of the parameters are $t_1=1.2,t_{-1}=0.5,\Delta_1=0.3$, and $\Delta_{-1}=-0.7$, with (a) $t_0=0.4$ and (b) $t_0=1.1$. ${\rm A}_i$ and ${\rm B}_i~(i=1,\cdots,4)$ are the cusps on the GBZ.}
\end{figure}

From Figs.~\ref{fig1}(b)-\ref{fig1}(e), one can see the nontrivial structure in the continuum bands. Within a single band, the energy eigenvalues change along either the real or the imaginary axis to some extent, and then they abruptly split off the axes. We find that this behavior comes from a unique and remarkable property that the single band is divided into three regions as shown in Fig.~\ref{fig2}. In the first region, the energies are real, and the eigenenergies of a time-reversal pair $\left(\ket{\psi},{\cal T}\ket{\psi}^\ast\right)$ are degenerate, i.e., $E=E^\ast$, where $\ket{\psi}$ is an eigenstate of the Hamiltonian (\ref{eq1}). In the second region, the energies are pure imaginary, and a pair of states $\ket{\psi}$ and $\Gamma{\cal T}\ket{\psi}^\ast$ related by the sublattice-time-reversal symmetry (STS) is degenerate, i.e., $E=-E^\ast$. In the third region, the energies are complex, and neither the time-reversal pair nor the sublattice-time-reversal pair is degenerate. We call these three regions the TRS-unbroken region, the STS-unbroken region, and the TRS/STS-broken region, respectively. We note that the gapless point appears at the boundary between the TRS-unbroken and STS-unbroken regions in Fig.~\ref{fig2}(b2).

%
%

\section{Mechanism for appearance of topological semimetal phase\label{sec3}}
In this section, we study the appearance of the gapless phase as shown in Fig.~\ref{fig1}(a) in terms of the non-Bloch band theory. First we describe this theory in the non-Hermitian SSH model (\ref{eq1}). The Bloch Hamiltonian of the non-Hermitian SSH model is expressed as an off-diagonal form 
\begin{eqnarray}
{\cal H}\left(\beta\right)=\left( \begin{array}{cc}
0                     & R_+\left(\beta\right) \\
R_-\left(\beta\right) & 0 \\
\end{array}\right),
\label{eq2}
\end{eqnarray}
with $R_\pm\left(\beta\right)=\left(t_1\pm\Delta_1\right)\beta\mp t_0+\left(t_{-1}\mp\Delta_{-1}\right)\beta^{-1}$, where $\beta\equiv{\rm e}^{ik},~k\in{\mathbb C}$. Then one can get the eigenvalue equation as $\det\left[{\cal H}\left(\beta\right)-E\right]=E^2-R_+\left(\beta\right)R_-\left(\beta\right)=0$, leading to the eigenenergies $E_\pm\left(\beta\right)$ with $E_-\left(\beta\right)=-E_+\left(\beta\right)$. From Ref.~\onlinecite{Yokomizo2019}, the condition for continuum bands is given by
\begin{equation}
\left|\beta_2\right|=\left|\beta_3\right|
\label{eq3}
\end{equation}
for four solutions of $\det\left[{\cal H}\left(\beta\right)-E\right]=0$ satisfying $\left|\beta_1\right|\leq\left|\beta_2\right|\leq\left|\beta_3\right|\leq\left|\beta_4\right|$, and the trajectories of $\beta_2$ and $\beta_3$ give the GBZ $C_\beta$. It is worth noting that $C_\beta$ is a closed loop encircling the origin on the complex plane~\cite{ZYang2019,KZhang2020}.

Next we describe the topological phase in the non-Hermitian SSH model (\ref{eq1}). The insulator phases of this system are classified in terms of a $Z$ topological invariant called winding number $w$ due to the SLS, defined as~\cite{Yokomizo2019}
\begin{equation}
w=-\frac{w_+-w_-}{2},~w_\pm=\frac{1}{2\pi}\left[\arg R_\pm\left(\beta\right)\right]_{C_\beta},
\label{eq4}
\end{equation}
where $\left[\arg R_\pm\left(\beta\right)\right]_{C_\beta}$ means the change of the phase of the functions $R_\pm\left(\beta\right)$ as $\beta$ goes along $C_\beta$ in a counterclockwise way. As long as there is a gap at $E=0$, $R_\pm\left(\beta\right)$ never vanish along $C_\beta$, and $w$ is well defined. Here, the gapped phase is defined as $E_+\left(\beta\right)\neq E_-\left(\beta\right)$ for every $\beta$ on $C_\beta$. In fact, Fig.~\ref{fig3}(a) shows the phase diagram, with the NI phase with $w=0$ (white region), the TI phase with $w=1$ (blue region), and the TSM phase (orange region). In the TSM phase, the gap closes at $E=0$, meaning that the equation $R_+\left(\beta\right)=0$ or $R_-\left(\beta\right)=0$ holds somewhere on $C_\beta$. At such a point on $C_\beta$, the Hamiltonian cannot be diagonalizable, and such point is called an exceptional point. Namely, in the TSM phase, the system has the exceptional points. Furthermore, the TSM phase appears as an intermediate phase between the NI and TI phases.

Now we explain the mechanism for the appearance of this TSM phase. In our model, the solutions of the equations $R_\pm\left(\beta\right)=0$ are gap-closing points, shown as the red and blue dots and squares in Figs.~\ref{fig3}(c1)-\ref{fig3}(h1), and they become exceptional points when $C_\beta$ goes through them. Let $\beta=\beta_i^a~(i=1,2,~a=+,-)$ denote the gap-closing points of $R_{a}\left(\beta\right)=0$, with $\left|\beta_1^a\right|\leq\left|\beta_2^a\right|$. In regions A and B in Fig.~\ref{fig3}(a), $\left|\beta_1^+\right|\leq\left|\beta_1^-\right|=\left|\beta_2^-\right|\leq\left|\beta_2^+\right|$ holds. Here the equality $\left|\beta_1^-\right|=\left|\beta_2^-\right|$ follows from $\beta_1^-=\left(\beta_2^-\right)^\ast$ because $\beta_1^-$ and $\beta_2^-$ are solutions of the algebraic equation $R_-\left(\beta\right)=0$ with real coefficients. Thus the condition (\ref{eq3}) is satisfied, and therefore $\beta_1^-$ and $\beta_2^-$ are on $C_\beta$. Hence they are the exceptional points. The condition $\beta_1^-=\left(\beta_2^-\right)^\ast$ remains satisfied even when the system parameter changes. Thus the exceptional points move along $C_\beta$ as shown in Figs.~\ref{fig3}(c2)-\ref{fig3}(h2), and the system remains in the TSM phase. In regions C and D in Fig.~\ref{fig3}(a), a similar scenario holds true by exchanging $R_+\left(\beta\right)$ and $R_-\left(\beta\right)$.

%
%

\section{Topological semimetal phase with exceptional points\label{sec4}}
In this section, we study in detail the TSM phase with exceptional points. Throughout the discussion, we can relate the creation and annihilation of the exceptional points with the change of the value of the winding number. Finally, we discuss the structure of the continuum band and of the GBZ in the TSM phase.

%
%

\subsection{Creation and annihilation of exceptional point\label{sec4-1}}
\begin{figure*}[]
\includegraphics[width=17.5cm]{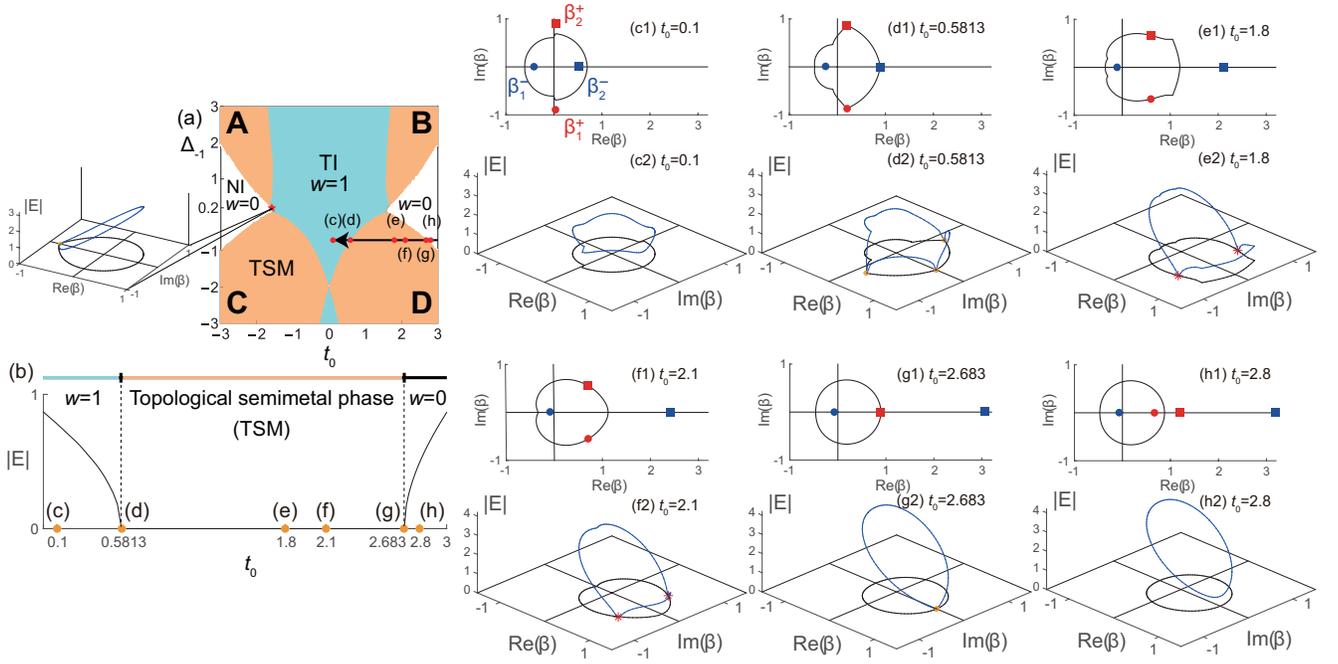}
\caption{\label{fig3}(a) Phase diagram in the non-Hermitian Su-Schrieffer-Heeger model for a long open chain with parameter values $t_1=1.2,t_{-1}=0.5$, and $\Delta_1=0.3$. At the red star ($t_0=-1.5922$ and $\Delta_{-1}=0.2$), the gap closes, and a direct transition between two insulator phases with $w=0$ (NI, white region) and $w=1$ (TI, blue region) occurs. The orange regions express the topological semimetal (TSM) phase. (b) Band gap along the black arrow ($\Delta_{-1}=-0.7$) in (a). Since it is a two-band model with the sublattice symmetry, the two eigenenergies satisfy $E_+\left(\beta\right)=-E_-\left(\beta\right)$, and we only show $\left|E\right|$ to see whether the gap closes. (c)-(h) Gap-closing points, generalized Brillouin zone, and motion of the exceptional points (red stars) along the black arrow in (a). The red (or blue) dots and squares express the gap-closing points of the equation $R_+\left(\beta\right)=0$ [or $R_-\left(\beta\right)=0$].}
\end{figure*}
In this section, we explain the creation and annihilation of the exceptional points in the non-Hermitian SSH model (\ref{eq1}). In the topological phase transition along the black arrow in Fig.~\ref{fig3}(a), we show the position of the gap-closing points $\beta_1^+$ and $\beta_2^+$ of the equation $R_+\left(\beta\right)=0$ [or $\beta_1^-$ and $\beta_2^-$ of $R_-\left(\beta\right)=0$] as the red (or blue) dots and squares, respectively, as shown in Figs.~\ref{fig3}(c1)-\ref{fig3}(h1). When we decrease the value of the parameter $t_0$, at $t_0=2.683$ [Fig.~\ref{fig3}(g-1)], $\beta_1^+$ and $\beta_2^+$ change from real values to complex values via coalescence, and this corresponds to a pair creation of exceptional points. After the coalescence, $\beta_1^+$ and $\beta_2^+$ become complex, with $\beta_1^+=\left(\beta_2^+\right)^\ast$, and their common absolute value is between the values of $\left|\beta_1^-\right|$ and $\left|\beta_2^-\right|$, meaning that $\beta_1^+$ and $\beta_2^+$ stay on the GBZ $C_\beta$. Then, at $t_0=0.5813$ [Fig.~\ref{fig3}(d1)], the value of $\left|\beta_2^-\right|$ becomes equal to that of $\left|\beta_1^+\right|\left(=\left|\beta_2^+\right|\right)$, and after passing that point (i.e., $t_0$ becomes less than 0.5813), $\left|\beta_1^-\right|<\left|\beta_2^-\right|<\left|\beta_1^+\right|=\left|\beta_2^+\right|$, meaning that $\beta_1^+$ and $\beta_2^+$ are no longer on $C_\beta$, and the gap opens [Fig.~\ref{fig3}(c1)]. At the phase-transition point $t_0=0.5813$ [Fig.~\ref{fig3}(d1)], three gap-closing points $\beta_1^+,\beta_2^+,\beta_2^-$ share the same absolute value, the gap closes at three points on $C_\beta$, and at such points, $C_\beta$ has cusps.

%
%

\subsection{Winding number\label{sec4-2}}
The change of the value of the winding number $w$ defined in Eq.~(\ref{eq4}) readily follows from the following argument. In the NI phase with $w=0$, $C_\beta$ surrounds one gap-closing point of $R_+\left(\beta\right)=0$ and one gap-closing point of $R_-\left(\beta\right)=0$ [Fig.~\ref{fig3}(h1)]. In this case, $w_\pm=\left[\arg R_\pm\left(\beta\right)\right]_{C_\beta}/2\pi=0$ because $R_\pm\left(\beta\right)$ is proportional to $\left(\beta-\beta_1^\pm\right)\left(\beta-\beta_2^\pm\right)/\beta$. On the other hand, in the TI phase with $w=1$, there exist two gap-closing points of $R_-\left(\beta\right)=0$ and no gap-closing points of $R_+\left(\beta\right)=0$ inside $C_\beta$ [Fig.~\ref{fig3}(c1)], which leads to $w_-=1$, $w_+=-1$, and $w=1$ as expected. Therefore the creation and annihilation of the exceptional points change the number of gap-closing points of $R_\pm\left(\beta\right)=0$ inside $C_\beta$, and the value of the topological invariant also changes.

We note that in addition to the topological phase transition between two insulator phases via the TSM phase with exceptional points, a direct phase transition from the NI phase to the TI phase is also possible as shown in the inset of Fig.~\ref{fig3}(a). Here, the gap closes on the real axis, where $R_+\left(\beta\right)$ and $R_-\left(\beta\right)$ simultaneously become zero at this value of $\beta$.

%
%

\subsection{TRS-unbroken phase, STS-unbroken phase, and TRS/STS-broken phase\label{sec4-3}}
Next we discuss the division of a single band into the TRS-unbroken region, the STS-unbroken region, and the TRS/STS-broken region, where the continuum band takes real, pure-imaginary, and general complex values, respectively. These regions are connected to each other at the cusps or the exceptional points. In Figs.~\ref{fig2}(a2) and \ref{fig2}(b2), three curves meet at one point, where $C_\beta$ have cusps represented by ${\rm A}_i$ and ${\rm B}_i~(i=1,\cdots,4)$. This comes from the property that three points on $C_\beta$ share the same energy. Furthermore, in Fig.~\ref{fig2}(b), the green and blue lines are connected at the exceptional point with $E=0$. Thus the exceptional point connecting the real and pure-imaginary energies becomes stable because such structure is topologically protected by the symmetries.

%
%

\section{General cases\label{sec5}}
In this section, we discuss the TSM phase with exceptional points in general cases. In this case, we show the mechanism for the appearance of the TSM phase and for the change of the winding number through the creation and annihilation of the exceptional points. Furthermore we discuss how the symmetry breaking affects the TSM phase.

%
%

\subsection{Topological semimetal phase and exceptional point\label{sec5-1}}
We show that the TSM phase appears in general cases. Due to the SLS, one can write the Bloch Hamiltonian ${\cal H}\left(\beta\right)$ as the off-diagonal form (\ref{eq2}) with $2N$ bands with $R_\pm\left(\beta\right)$ being $N\times N$ matrices. Then the eigenvalue equation $\det\left[{\cal H}\left(\beta\right)-E\right]=0$ yields bands symmetric with respect to $E=0$. Furthermore the condition for continuum bands is given by $\left|\beta_M\right|=\left|\beta_{M+1}\right|$ for the solutions $\beta_1,\cdots,\beta_{2M}\left(\left|\beta_1\right|\leq\cdots\leq\left|\beta_{2M}\right|\right)$ of $\det\left[{\cal H}\left(\beta\right)-E\right]=0$, being an algebraic equation for $\beta$ with an even degree $2M$. Now a condition for a gap closing at $E=0$ is decomposed into two equations $\det R_+\left(\beta\right)=0$ and $\det R_-\left(\beta\right)=0$. Thanks to the TRS, $\det R_\pm\left(\beta\right)$ are polynomials of $\beta$ and $\beta^{-1}$ with real coefficients. Namely, it follows that any complex solutions of $\det R_\pm\left(\beta\right)=0$ appear in complex conjugate pairs $(\beta,\beta^\ast)$. Then, when we suppose $\beta_M$ and $\beta_{M+1}$ form a pair of the complex conjugate solutions of $\det R_+\left(\beta\right)=0$ (or $\det R_-\left(\beta\right)=0$), we have $\left|\beta_M\right|=\left|\beta_{M+1}\right|$, meaning that $\beta_M$ and $\beta_{M+1}$ are on the GBZ $C_\beta$, and the gap closes. Therefore, even when system parameters change, the gap remains zero as long as this pair gives the $M$th and $(M+1)$th largest absolute values among the $2M$ solutions.

\begin{figure}[]
\includegraphics[width=8.5cm]{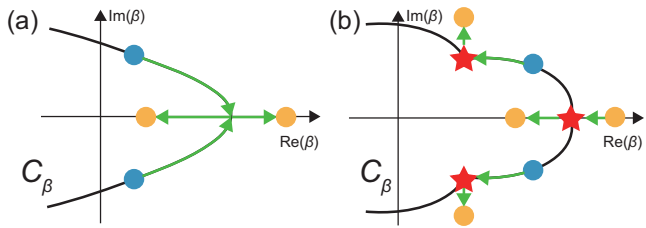}
\caption{\label{fig4}(a) Coalescence of the exceptional points. (b) Annihilation of the exceptional point at the cusp. The orange and blue dots express the gap-closing points and the exceptional points, respectively, and the red stars are the cusps of the generalized Brillouin zone $C_\beta$.}
\end{figure}

Then, because of the mechanism for the stabilization of the exceptional points as discussed in Sec.~\ref{sec4}, annihilations (and likewise creations) of them are limited to two patterns as shown in Figs.~\ref{fig4}(a) and \ref{fig4}(b). Figure~\ref{fig4}(a) represents a coalescence of two exceptional points, and Fig.~\ref{fig4}(b) represents an encounter between the gap-closing point and the cusp. In Fig.~\ref{fig4}(a), the two exceptional points meet, become two real gap-closing points, and move away from the GBZ $C_\beta$. This can be seen in Fig.~\ref{fig3}(g). On the other hand, the case of Fig.~\ref{fig4}(b) occurs when two complex-conjugate exceptional points and one gap-closing point share the same absolute value. If this occurs, for example in Fig.~\ref{fig3}(d), where $\left|\beta_1^-\right|=\left|\beta_2^-\right|=\left|\beta_2^+\right|$ is satisfied, the gap closes at the three points on $C_\beta$. As we change the system parameters, the ordering of the absolute values of three gap-closing points $\beta_{1,2}^-$ and $\beta_2^+$ changes, allowing the exceptional point to disappear and the gap to open.

%
%

\subsection{Change of winding number\label{sec5-2}}
In this section, we show that in one-dimensional (1D) non-Hermitian systems with both SLS and TRS, the value of the winding number changes through the creation and annihilation of the exceptional points. In the following, we focus on a two-band model. For the Bloch Hamiltonian ${\cal H}\left(\beta\right)$ written as the off-diagonal form, i.e., Eq.~(\ref{eq2}), without loss of generality, we can write the holomorphic functions $R_\pm\left(\beta\right)$ as
\begin{equation}
R_\pm\left(\beta\right)=\frac{C_\pm}{\beta^m}\prod_{i=1}^{2m}\left(\beta-\beta_i^\pm\right),
\label{eq5}
\end{equation}
where $C_\pm$ are real constants due to the TRS. Then $\det\left[{\cal H}\left(\beta\right)-E\right]=R_+\left(\beta\right)R_-\left(\beta\right)-E^2=0$ can be explicitly written as
\begin{equation}
\frac{C_+C_-}{\beta^{2m}}\prod_{i=1}^{2m}\left(\beta-\beta_i^+\right)\left(\beta-\beta_i^-\right)=E^2,
\label{eq6}
\end{equation}
which is an algebraic equation for $\beta$ with a degree $4m$. Here, by numbering the solutions of Eq.~(\ref{eq6}) so as to satisfy $\left|\beta_1\right|\leq\cdots\leq\left|\beta_{4m}\right|$, the condition for continuum bands is given by
\begin{equation}
\left|\beta_{2m}\right|=\left|\beta_{2m+1}\right|,
\label{eq7}
\end{equation}
and one can get $C_\beta$ from the trajectories of $\beta_{2m}$ and $\beta_{2m+1}$.

From Eq.~(\ref{eq5}), we can rewrite Eq.~(\ref{eq4}) as
\begin{equation}
w=-\frac{N_{\rm zeros}^+-N_{\rm zeros}^-}{2},
\label{eq8}
\end{equation}
where $N_{\rm zeros}^\pm$ expresses the number of the solutions $\beta_i^\pm$ of $R_\pm\left(\beta\right)=0$ inside $C_\beta$, respectively. Furthermore, it is worth noting that from Ref.~\onlinecite{KZhang2020}, when the system has a gap around $E=0$, we can get
\begin{equation}
\frac{1}{2\pi}\int_{C_\beta}{\rm d}\log\det{\cal H}\left(\beta\right)=N_{\rm zeros}-2m=0,
\label{eq9}
\end{equation}
where $N_{\rm zeros}\left(=N_{\rm zero}^++N_{\rm zero}^-\right)$ expresses the number of solutions of the equation $\det{\cal H}\left(\beta\right)=0$ inside $C_\beta$. Equation~(\ref{eq9}) tells us that the total number of the solutions of $\det{\cal H}\left(\beta\right)=0$ inside $C_\beta$, namely, the number of the gap-closing points inside $C_\beta$, is unchanged as long as the system has a gap.

Now we focus on two insulator phases separated by the TSM phase with exceptional points. When the system enters the TSM phase from one of the insulator phases, the exceptional points are created by the inverse process as shown in Fig.~\ref{fig4}(a) or \ref{fig4}(b). Then, after a further change of the system parameters, the system becomes the other insulator phase from the TSM phase, and here, the exceptional points are annihilated by the process as shown in Fig.~\ref{fig4}(a) or \ref{fig4}(b). If the creation is by the inverse process of Fig.~\ref{fig4}(a) and the annihilation is by the process of Fig.~\ref{fig4}(b) (or vice versa), while the total number $N_{\rm zeros}\left(=N_{\rm zero}^++N_{\rm zero}^-\right)$ of the solutions of $\det{\cal H}\left(\beta\right)=0$ inside $C_\beta$ is unchanged, such a motion of the exceptional points changes the value of $N_{\rm zeros}^+$ and $N_{\rm zeros}^-$ by $1$ and $-1$ (or by $-1$ and $1$), respectively, resulting in the change of the value of $w$ by $1$ (or by $-1$). Therefore we conclude that these insulator phases have different values of $w$. We note that this scenario can be extended to multiband systems.

%
%

\subsection{Symmetry-breaking effect\label{sec5-3}}
Here, we discuss the effect of symmetry breaking. When the SLS is broken while the TRS is preserved, the gapless phase survives because of the reality of the coefficients of $\det\left[{\cal H}\left(\beta\right)-E\right]=0$, but the gap closes not necessarily at $E=0$. On the other hand, when the TRS is broken but either the SLS or the pseudo-particle-hole-symmetry (PHS)~\cite{Kawabata2019v3} is preserved, the TSM phase disappears (see Appendix~\ref{secA}) because the coefficients of $\det\left[{\cal H}\left(\beta\right)-E\right]=0$ become complex. In conclusion, the TSM phase in 1D non-Hermitian systems is robust against the change of system parameters, provided both the SLS and the TRS are preserved.

%
%

\section{Summary\label{sec6}}
In summary, the appearance of the TSM phase with exceptional points is attributed to the unique features of the GBZ $C_\beta$. It is shown that in 1D non-Hermitian systems with the SLS and TRS, the TSM phase is stable, unlike Hermitian systems. We also find that each energy band is divided into three regions in terms of the symmetry of the eigenstates, and the regions switch only at the cusps and the exceptional points on $C_\beta$. Thus non-Hermiticity brings about qualitative changes to the topological phase transition and internal structure of energy bands.

We should comment on the experimental realization of the TSM phase. A simple case of the non-Hermitian SSH model was experimentally realized in an electric circuit~\cite{Helbig2020}. Therefore we expect that our theory can be verified by adding next-nearest-neighbor hopping terms. Since the only restrictions are the SLS and TRS, there remains much room for the choice of parameter values, toward experimental verification.

%
%

\begin{acknowledgments}
We are grateful to Ryo Okugawa and Ryo Takahashi for valuable discussions. This work was supported by JSPS KAKENHI Grant No.~JP18H03678 and by MEXT Elements Strategy Initiative to Form Core Research Centers Grant No.~JPMXP0112101001. K. Y. was also supported by JSPS KAKENHI Grant No.~JP18J22113.
\end{acknowledgments}

%
%

\appendix

%
%

\section{ONE-DIMENSIONAL NON-HERMITIAN SYSTEM WITH PSEUDO-particle-hole symmetry\label{secA}}
In this appendix, we investigate a 1D non-Hermitian system with pseudo-particle-hole-symmetry (PHS)~\cite{Kawabata2019v3} and show that it is classified in terms of a $Z_2$ topological invariant. Here, for a real-space Hamiltonian $H$, this symmetry is defined as
\begin{equation}
{\cal C}H^\ast{\cal C}^{-1}=-H,
\label{eq10a}
\end{equation}
where ${\cal C}$ is the unitary matrix, and ${\cal C}{\cal C}^\ast=+1$. This pseudo-PHS is the product between the SLS and the TRS. In the following, we focus on a two-band model.

%
%

\subsection{Two-band model\label{secA-1}}
The Bloch Hamiltonian ${\cal H}\left(\beta\right)$ can be written as
\begin{equation}
{\cal H}\left(\beta\right)={\cal H}_0\left(\beta\right)\sigma_0+\sum_{i=x,y,z}{\cal H}_i\left(\beta\right)\sigma_i,
\label{eq11a}
\end{equation}
where $\sigma_0$ is a $2\times2$ identity matrix, $\sigma_i~\left(i=x,y,z\right)$ are the Pauli matrices, and the complex Bloch wave number is defined as $\beta\equiv{\rm e}^{ik},~k\in{\mathbb C}$. We assume that it satisfies
\begin{equation}
\sigma_x\left[{\cal H}\left(\beta\right)\right]^\ast\sigma_x^{-1}=-{\cal H}\left(\beta^\ast\right),
\label{eq12a}
\end{equation}
and then, the coefficients ${\cal H}_i\left(\beta\right)~\left(i=0,x,y,z\right)$ in Eq.~(\ref{eq11a}) satisfy
\begin{eqnarray}
\left[{\cal H}_i\left(\beta\right)\right]^\ast=-{\cal H}_i\left(\beta^\ast\right)~\left(i=0,x,y\right),~\left[{\cal H}_z\left(\beta\right)\right]^\ast={\cal H}_z\left(\beta^\ast\right). \nonumber\\
\label{eq13a}
\end{eqnarray}
It is worth noting that ${\cal H}_i\left(\beta\right)~\left(i=0,x,y\right)$ are pure imaginary and ${\cal H}_z\left(\beta\right)$ is real when $\arg\beta=0$ and $\arg\beta=\pi$.

%
%

\subsection{$Z_2$ topological invariant\label{secA-2}}
Such a system is classified in terms of the $Z_2$ topological invariant $\nu~(=0,1)$, and for the Bloch Hamiltonian (\ref{eq11a}), we can define it as
\begin{eqnarray}
&&\nu=\frac{1}{2\pi}\int_{\beta_0}^{\beta_\pi}{\rm d}\beta\frac{\rm d}{{\rm d}\beta}\left[\arg{\cal R}_+\left(\beta\right)-\arg{\cal R}_-\left(\beta\right)\right]~\left({\rm mod}~2\right), \nonumber\\
&&{\cal R}_\pm\left(\beta\right)={\cal H}_z\left(\beta\right)\pm i\sqrt{{\cal H}_x^2\left(\beta\right)+{\cal H}_y^2\left(\beta\right)},
\label{eq11b}
\end{eqnarray}
where $\beta_0$ (or $\beta_\pi$) is the value of $\beta$ at $\arg\beta=0$ (or $\arg\beta=\pi$) on the GBZ $C_\beta$ (for example, see Fig.~\ref{figS2}). In Eq.~(\ref{eq11b}), the integral contour $\beta$ goes along $C_\beta$, and we select the branch cut of the square root so that both functions ${\cal R}_\pm\left(\beta\right)$ become continuous on $C_\beta$.

In the following, we assume that a system has a gap. Here, we note that two continuum bands are separated by a line which determines a complex gap on the complex energy plane. This gap is called a line gap~\cite{Kawabata2019}. In this case, one can show that $\nu$ takes only $0$ or $1$. To this end, we calculate the value of $\exp\left(2\pi i\nu\right)$. As mentioned in Appendix~\ref{secA-1}, since the functions ${\cal R}_\pm\left(\beta_0\right)$ and ${\cal R}_\pm\left(\beta_\pi\right)$ take real values, we can rewrite the expression of $\exp\left(2\pi i\nu\right)$ as
\begin{eqnarray}
\exp\left(2\pi i\nu\right)&=&\exp\left\{i\left[\arg{\cal R}_+\left(\beta_\pi\right)-\arg{\cal R}_+\left(\beta_0\right)\right]\right\} \nonumber\\
&\times&\exp\left\{i\left[\arg{\cal R}_-\left(\beta_\pi\right)-\arg{\cal R}_-\left(\beta_0\right)\right]\right\} \nonumber\\
&=&\frac{{\rm sgn}\left[{\cal R}_+\left(\beta_\pi\right)\right]}{{\rm sgn}\left[{\cal R}_+\left(\beta_0\right)\right]}\frac{{\rm sgn}\left[{\cal R}_-\left(\beta_\pi\right)\right]}{{\rm sgn}\left[{\cal R}_-\left(\beta_0\right)\right]} \nonumber\\
&=&\prod_{\beta=\beta_0,\beta_\pi}\prod_{\sigma=\pm}{\rm sgn}\left[{\cal R}_\sigma\left(\beta\right)\right] \nonumber\\
&=&\prod_{\beta=\beta_0,\beta_\pi}{\rm sgn}\left[\sum_{i=x,y,z}{\cal H}_i^2\left(\beta\right)\right]. \nonumber\\
\label{eq12b}
\end{eqnarray}
Here we note that the quantities $\sum_i{\cal H}_i^2\left(\beta_0\right)$ and $\sum_i{\cal H}_i^2\left(\beta_\pi\right)$ are real. Since we assume the presence of a line gap, we conclude that $\sum_i{\cal H}_i^2\left(\beta_0\right)$ and $\sum_i{\cal H}_i^2\left(\beta_\pi\right)$ have the same sign, as we prove by contradiction in the following.

Suppose $\sum_i{\cal H}_i^2\left(\beta_0\right)$ and $\sum_i{\cal H}_i^2\left(\beta_\pi\right)$ have different signs. We can set
\begin{equation}
\sum_{i=x,y,z}{\cal H}_i^2\left(\beta_0\right)>0,~\sum_{i=x,y,z}{\cal H}_i^2\left(\beta_\pi\right)<0
\label{eq13b}
\end{equation}
without loss of generality. First of all, we assume ${\cal H}_0\left(\beta\right)=0$ for simplicity. At $\beta=\beta_0$, the energies are $E=\pm\varepsilon_0$, $\varepsilon_0=\sqrt{\sum_i{\cal H}_i^2\left(\beta_0\right)}>0$. Now we choose $E=\varepsilon_0=\sqrt{\sum_i{\cal H}_i^2\left(\beta_0\right)}$, and we change $\beta$ along $C_\beta$ in a counterclockwise way from $\beta=\beta_0$ to $\beta=\beta_\pi$. Here, let $C_+$ denote this path on the complex plane. Then, at $\beta=\beta_\pi$, the energy is given by $E=\varepsilon_\pi=\sqrt{\sum_i{\cal H}_i^2\left(\beta_\pi\right)}\in i{\mathbb R}$, where the branch of the square root is chosen in such a way that $E=\sqrt{\sum_i{\cal H}_i^2\left(\beta\right)}$ is continuous along $C_+$. Next we consider a path $C_-$ along $C_\beta$ in a clockwise way from $\beta=\beta_0$ to $\beta=\beta_\pi$. Because $\left[\sum_i{\cal H}_i^2\left(\beta\right)\right]^\ast=\left[\sum_i{\cal H}_i^2\left(\beta^\ast\right)\right]$, the energy at $\beta$ and that at $\beta^\ast$ are complex conjugate. Therefore, because $\left(C_+\right)^\ast=C_-$, the energy at $\beta=\beta_\pi$ along $C_-$ is $E=\varepsilon_\pi^\ast=-\varepsilon_\pi$. Thus, by encircling $C_\beta\left(=-C_-+C_+\right)$ once from $\beta_\pi$ to $\beta_\pi$, the branch changes from $E=-\varepsilon_\pi$ to $E=\varepsilon_\pi$, meaning that the two energies $E=\pm\sqrt{\sum_i{\cal H}_i^2\left(\beta\right)}$ are continuously connected, and there is no line gap, contradicting the assumption. Thus we can conclude that $\sum_i{\cal H}_i^2\left(\beta_0\right)$ and $\sum_i{\cal H}_i^2\left(\beta_\pi\right)$ have the same sign. So far, we assume ${\cal H}_0\left(\beta\right)=0$, but even in the case of ${\cal H}_0\left(\beta\right)\neq0$, because this term does not affect the above argument, the above proof remains valid.

Therefore, from Eq.~(\ref{eq12b}), we can get $\exp\left(2\pi i\nu\right)=1$ and can conclude that the value of $\nu$ is an integer. As a result, $\nu$ can take only $0$ or $1$ (mod $2$), and we conclude that $\nu$ can be interpreted as a $Z_2$ topological invariant in this system.

In particular, in Hermitian cases, we can greatly simplify the formula of $\nu$ in Eq.~(\ref{eq11b}). The Bloch wave number $k$ becomes real, and all the coefficients included in Eq.~(\ref{eq11a}) become real functions. In the following, we replace ${\cal H}_i\left(\beta\right)$ by $H_i\left(k\right)~\left(i=0,x,y,z\right),~k\in\left[-\pi,\pi\right]$. Here, the system has the conventional PHS represented as
\begin{eqnarray}
H_i\left(k\right)=-H_i\left(-k\right)~\left(i=0,x,y\right),~H_z\left(k\right)=H_z\left(-k\right), \nonumber\\
\label{eq14b}
\end{eqnarray}
and $H_i\left(k\right)~\left(i=0,x,y\right)$ become zero at $k=0$ and $k=\pi$. Furthermore, we can get
\begin{eqnarray}
\arg\left(H_z-i\sqrt{H_x^2+H_y^2}\right)=-\arg\left(H_z+i\sqrt{H_x^2+H_y^2}\right), \nonumber\\
\label{eq15b}
\end{eqnarray}
and then, Eq.~(\ref{eq11b}) can be rewritten as
\begin{eqnarray}
\nu&=&\frac{1}{\pi}\int_{0}^\pi{\rm d}k\frac{\rm d}{{\rm d}k}\arg\left[H_z\left(k\right)+i\sqrt{H_x^2\left(k\right)+H_y^2\left(k\right)}\right] \nonumber\\
&=&\frac{1}{\pi}\left[\arg H_z\left(k\right)\right]_0^\pi.
\label{eq16b}
\end{eqnarray}
Since Eq.~(\ref{eq14b}) tells us that both ${\rm arg}H_z\left(0\right)$ and ${\rm arg}H_z\left(\pi\right)$ take $0$ or $\pi$ (mod $2\pi$), Eq.~(\ref{eq16b}) can be further rewritten as
\begin{eqnarray}
\nu=\left\{ \begin{array}{ll}
0 & {\rm if}~{\rm sgn}\left[H_z\left(0\right)\right]{\rm sgn}\left[H_z\left(\pi\right)\right]>0, \vspace{5pt}\\
1 & {\rm if}~{\rm sgn}\left[H_z\left(0\right)\right]{\rm sgn}\left[H_z\left(\pi\right)\right]<0,
\end{array}\right.
\label{eq17b}
\end{eqnarray}
and we obtain the known formula~\cite{Kitaev2001}
\begin{equation}
\left(-1\right)^\nu={\rm sgn}\left[H_z\left(0\right)\right]{\rm sgn}\left[H_z\left(\pi\right)\right].
\label{eq18b}
\end{equation}

%
%

\subsection{Generalized non-Hermitian Su-Schrieffer-Heeger model\label{secA-3}}
\begin{figure}[]
\includegraphics[width=8.5cm]{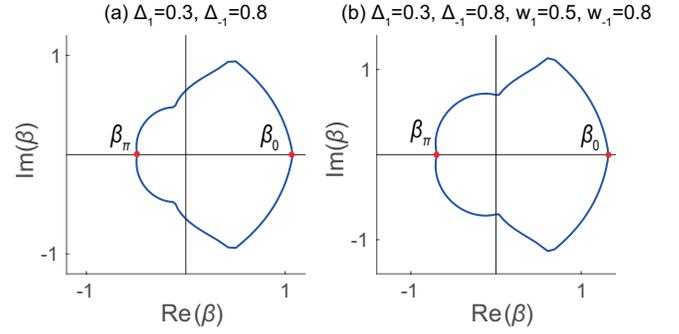}
\caption{\label{figS2}Generalized Brillouin zone $C_\beta$ in the generalized non-Hermitian Su-Schrieffer-Heeger model. The values of the parameters are $t=t^\prime=0,t_1=1.2,t_0=1$, and $t_{-1}=0.5$; (a) $\Delta_1=0.3,\Delta_{-1}=0.8$, and $w_1=w_{-1}=0$, and (b) $\Delta_1=0.3,\Delta_{-1}=0.8,w_1=0.5$, and $w_{-1}=0.8$. The points where $C_\beta$ intersects the positive and negative sides of a real axis are denoted by $\beta_0$ and $\beta_\pi$, respectively.}
\end{figure}
In this section, we study the generalized non-Hermitian Su-Schrieffer-Heeger (SSH) model. The real-space Hamiltonian of this system can be written as
\begin{eqnarray}
H&=&\sum_n\left(it_1^{\rm A}c^\dag_{n,{\rm A}}c_{n+1,{\rm A}}+it_0^{\rm A}c^\dag_{n,{\rm A}}c_{n,{\rm A}}+it_{-1}^{\rm A}c^\dag_{n+1,{\rm A}}c_{n,{\rm A}}\right. \nonumber\\
&+&it_1^{\rm B}c^\dag_{n,{\rm B}}c_{n+1,{\rm B}}+it_0^{\rm B}c^\dag_{n,{\rm B}}c_{n,{\rm B}}+it_{-1}^{\rm B}c^\dag_{n+1,{\rm B}}c_{n,{\rm B}} \nonumber\\
&+&t_1^+c_{n,{\rm A}}^\dag c_{n+1,{\rm B}}+t_0^+c^\dag_{n,{\rm A}}c_{n,{\rm B}}+t_{-1}^+c_{n+1,{\rm A}}^\dag c_{n,{\rm B}} \nonumber\\
&+&\left.t_1^-c_{n,{\rm B}}^\dag c_{n+1,{\rm A}}+t_0^-c^\dag_{n,{\rm B}}c_{n,{\rm A}}+t_{-1}^-c_{n+1,{\rm B}}^\dag c_{n,{\rm A}}\right),
\label{eq11c}
\end{eqnarray}
where all the parameters are set to be real, meaning that the real-space Hamiltonian (\ref{eq11c}) satisfies the pseudo PHS. Henceforth we set all the parameters as $t_0^{\rm A}=t_0^{\rm B}=t$, $t_{\pm1}^{\rm A}=t^\prime\mp w_{\pm1}$, $t_{\pm}^{\rm B}=t^\prime\pm w_{\pm1}$, $t_0^+=t_0^-=-t_0$, $t_1^\pm=t_1\pm\Delta_1$, and $t_{-1}^\pm=t_{-1}\mp\Delta_{-1}$. Here, for the Bloch Hamiltonian ${\cal H}\left(\beta\right)$ in the form (\ref{eq11a}), the coefficients ${\cal H}_i\left(\beta\right)~\left(i=0,x,y,z\right)$ are given by
\begin{eqnarray}
{\cal H}_0\left(\beta\right)&=&it+it^\prime\left(\beta+\beta^{-1}\right), \nonumber\\
{\cal H}_x\left(\beta\right)&=&t_1\beta-t_0+t_{-1}\beta^{-1}, \nonumber\\
{\cal H}_y\left(\beta\right)&=&i\left(\Delta_1\beta-\Delta_{-1}\beta^{-1}\right), \nonumber\\
{\cal H}_z\left(\beta\right)&=&-i\left(w_1\beta-w_{-1}\beta^{-1}\right).
\label{eq12c}
\end{eqnarray}
It satisfies the condition for the pseudo-PHS given by Eq.~(\ref{eq12a}) with the replacement $x\rightarrow y,y\rightarrow z$, and $z\rightarrow x$. As we discussed previously, it can be classified in terms of the $Z_2$ topological invariant $\nu$ defined in Eq.~(\ref{eq11b}). We note that in this case, the form of $\nu$ can be obtained as Eq.~({\ref{eq11b}}) by replacing the variables: $x\rightarrow y,y\rightarrow z$, and $z\rightarrow x$.

\begin{figure}[]
\includegraphics[width=8.5cm]{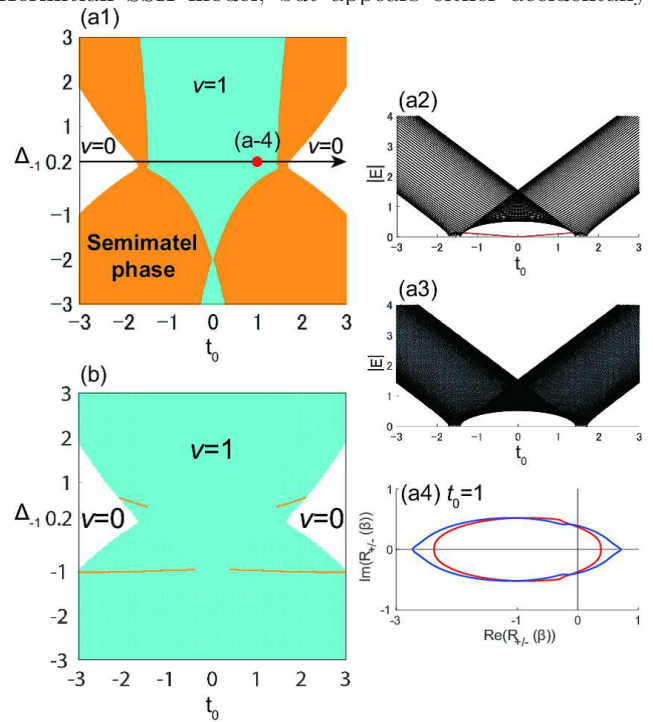}
\caption{\label{figS3}(a) Phase diagram and bulk-edge correspondence in the generalized non-Hermitian Su-Schrieffer-Heeger (SSH) model with $t=t^\prime=w_{-1}=0,t_1=1.2,t_{-1}=0.5,\Delta_1=0.3$, and $w_1=0.2$. We note that in this case, ${\cal H}_0\left(\beta\right)=0$. (a1) Phase diagram on the $t_0$-$\Delta_{-1}$ plane. The blue region represents a topological insulator phase with the $Z_2$ topological invariant $\nu$ equal to 1, the white region represents a normal insulator phase phase with $\nu=0$, and the orange region represents a semimetal phase with exceptional points. Along the black arrow in (a1) with $\Delta_{-1}=0.2$, we show the results for (a2) energy levels in a finite open chain and (a3) the continuum bands. The edge states are shown in red in (a2). (a4) shows $\ell_+$ (red) and $\ell_-$ (blue) on the ${\bm R}$ plane with $t_0=1$ and $\Delta_{-1}=0.2$. These loops encircle the origin on the complex plane, meaning that the value of $\nu$ is equal to $1$. (b) Phase diagram in the generalized non-Hermitian SSH model with ${\cal H}_0\left(\beta\right)\neq0$. The parameters are set as $t=w_{-1}=0,t_1=1.2,t_{-1}=0.5,\Delta_1=0.3$, and $w_1=0.2$, and the parameter $t^\prime$ takes infinitesimal values.}
\end{figure}

The eigenvalue equation $\det\left[{\cal H}\left(\beta\right)-E\right]=0$ is a quartic equation for $\beta$, the condition for continuum bands can be written as $\left|\beta_2\right|=\left|\beta_3\right|$ when the solutions satisfy $\left|\beta_1\right|\leq\left|\beta_2\right|\leq\left|\beta_3\right|\leq\left|\beta_4\right|$. We note that the trajectories of $\beta_2$ and $\beta_3$ give the GBZ $C_\beta$. Examples of $C_\beta$ and the continuum bands are given in Fig.~\ref{figS2} and in Fig.~\ref{figS3}(a3), respectively.

Let $\ell_\pm$ denote the loops drawn by the functions ${\cal R}_\pm\left(\beta\right)$ on the complex plane when $\beta$ goes along $C_\beta$ in a counterclockwise way. Equation~(\ref{eq11b}) tells us how to determine the value of $\nu$; when neither $\ell_+$ nor $\ell_-$ surrounds the origin ${\cal O}$ on the complex plane, $\nu$ is equal to 0, and when two loops simultaneously surround ${\cal O}$, $\nu$ is equal to 1. For example, in the case of Fig.~\ref{figS3}(a4), $\nu$ becomes 1. We note that the system has exceptional points when either $\ell_+$ or $\ell_-$ passes ${\cal O}$, and $\nu$ is not well defined in this case.

We can get the phase diagram in the generalized non-Hermitian SSH model as shown in Fig.~\ref{figS3}(a1) and can confirm that the topological edge states appear when $\nu$ takes the nonzero value as shown in Fig.~\ref{figS3}(a2). Therefore we can establish the bulk-edge correspondence between the $Z_2$ topological invariant $\nu$ and existence of the topological edge states.

In this model, we can see that the pseudo PHS cannot topologically protect this semimetal phase. For simplicity, let the value of the parameter $t$ be zero. So far, we have set $t=t^\prime=0$ in the calculations in Fig.~\ref{figS3}(a1)-\ref{figS3}(a4). On the other hand, for $t^\prime\neq0$ being an infinitesimal value, as an example, we can obtain the phase diagram as shown in Fig.~\ref{figS3}(b). We can confirm that the semimetal phase in Fig.~\ref{figS3}(a1) disappears by putting $t^\prime\neq0$. We note that the exceptional points appear on the orange lines on the phase diagram in Fig.~\ref{figS3}(b). However, these exceptional points can be removed by adding other perturbation terms. Therefore we conclude that the pseudo PHS cannot topologically protect the semimetal phase.

%
%

\section{PREVIOUS WORKS ON THE NON-HERMITIAN SSH MODEL\label{secB}}
The non-Hermitian SSH model, which is one of the simplest non-Hermitian models, has been studied in many previous works. In this appendix, we explain that these previous works did not deal with the important and unique properties of this model found in our work.

As mentioned in the main text, in general non-Hermitian systems, there is a large difference between the energy spectrum in a periodic chain and that in an open chain, unlike Hermitian systems. An open chain is of particular interest, because of its novel properties as well as its importance in real systems. Our main findings in the present paper are unique to an open chain showing the non-Hermitian skin effect due to the complex Bloch wave number. Therefore they cannot be studied through the previous works~\cite{Yin2018,Herviou2019,YWu2019,RChen2019,LLi2019,Loic2019,Wang2020} on the energy spectrum, the eigenstates, and the topological invariant in the non-Hermitian SSH model with periodic boundary conditions.

Furthermore, even in an open chain, the Bloch wave number becomes real in special cases of the non-Hermitian SSH model~\cite{Runder2009,Liang2013,Zhu2014,Zhao2015,Jin2017,Yuce2018pra,Lieu2018,Klett2018,Yuce2018,Ge2019,Fu2020,Poli2015,Zeuner2015,Weimann2017,St2017,Parto2018,Pan2018}. In such cases, the non-Hermitian skin effect and the properties of the system discovered in our work do not appear. We note that this real Bloch wave number in the previous works~\cite{Runder2009,Liang2013,Zhu2014,Zhao2015,Jin2017,Yuce2018pra,Lieu2018,Klett2018,Yuce2018,Ge2019,Fu2020,Poli2015,Zeuner2015,Weimann2017,St2017,Parto2018,Pan2018} does not come from the fundamental symmetries (i.e., the SLS and the TRS) of the non-Hermitian SSH model, but appears either accidentally or by adding the parity-time $({\cal P}{\cal T})$ symmetry.

Finally, we mention that in the previous works~\cite{Kunst2018,Yao2018,Jin2019,Lee2019,Kunst2019,Deng2019,Imura2019,Song2019,Song2019v2,Longhi2019,Ghatak2019e,Helbig2020}, the Bloch wave number becomes complex. In these previous works, the eigenstates localized at either end of an open chain were mainly focused on because these localized states can lead to the novel nonreciprocal phenomena. Nevertheless, the TSM phase studied in the present paper has not been found in these previous works. Our findings are on the absence/presence of the gap and the symmetry of the eigenstates, both of which are vital for understanding the physics of non-Hermitian systems.

%

\providecommand{\noopsort}[1]{}\providecommand{\singleletter}[1]{#1}%

\begin{thebibliography}{132}%
\makeatletter
\providecommand \@ifxundefined [1]{%
 \@ifx{#1\undefined}
}%
\providecommand \@ifnum [1]{%
 \ifnum #1\expandafter \@firstoftwo
 \else \expandafter \@secondoftwo
 \fi
}%
\providecommand \@ifx [1]{%
 \ifx #1\expandafter \@firstoftwo
 \else \expandafter \@secondoftwo
 \fi
}%
\providecommand \natexlab [1]{#1}%
\providecommand \enquote  [1]{``#1''}%
\providecommand \bibnamefont  [1]{#1}%
\providecommand \bibfnamefont [1]{#1}%
\providecommand \citenamefont [1]{#1}%
\providecommand \href@noop [0]{\@secondoftwo}%
\providecommand \href [0]{\begingroup \@sanitize@url \@href}%
\providecommand \@href[1]{\@@startlink{#1}\@@href}%
\providecommand \@@href[1]{\endgroup#1\@@endlink}%
\providecommand \@sanitize@url [0]{\catcode `\\12\catcode `\$12\catcode
  `\&12\catcode `\#12\catcode `\^12\catcode `\_12\catcode `\%12\relax}%
\providecommand \@@startlink[1]{}%
\providecommand \@@endlink[0]{}%
\providecommand \url  [0]{\begingroup\@sanitize@url \@url }%
\providecommand \@url [1]{\endgroup\@href {#1}{\urlprefix }}%
\providecommand \urlprefix  [0]{URL }%
\providecommand \Eprint [0]{\href }%
\providecommand \doibase [0]{http://dx.doi.org/}%
\providecommand \selectlanguage [0]{\@gobble}%
\providecommand \bibinfo  [0]{\@secondoftwo}%
\providecommand \bibfield  [0]{\@secondoftwo}%
\providecommand \translation [1]{[#1]}%
\providecommand \BibitemOpen [0]{}%
\providecommand \bibitemStop [0]{}%
\providecommand \bibitemNoStop [0]{.\EOS\space}%
\providecommand \EOS [0]{\spacefactor3000\relax}%
\providecommand \BibitemShut  [1]{\csname bibitem#1\endcsname}%
\let\auto@bib@innerbib\@empty
\bibitem [{\citenamefont {Eichelkraut}\ \emph {et~al.}(2013)\citenamefont
  {Eichelkraut}, \citenamefont {Heilmann}, \citenamefont {Weimann},
  \citenamefont {St{\"u}tzer}, \citenamefont {Dreisow}, \citenamefont
  {Christodoulides}, \citenamefont {Nolte},\ and\ \citenamefont
  {Szameit}}]{Eichelkraut2013}%
  \BibitemOpen
  \bibfield  {author} {\bibinfo {author} {\bibfnamefont {T.}~\bibnamefont
  {Eichelkraut}}, \bibinfo {author} {\bibfnamefont {R.}~\bibnamefont
  {Heilmann}}, \bibinfo {author} {\bibfnamefont {S.}~\bibnamefont {Weimann}},
  \bibinfo {author} {\bibfnamefont {S.}~\bibnamefont {St{\"u}tzer}}, \bibinfo
  {author} {\bibfnamefont {F.}~\bibnamefont {Dreisow}}, \bibinfo {author}
  {\bibfnamefont {D.}~\bibnamefont {Christodoulides}}, \bibinfo {author}
  {\bibfnamefont {S.}~\bibnamefont {Nolte}}, \ and\ \bibinfo {author}
  {\bibfnamefont {A.}~\bibnamefont {Szameit}},\ }\href@noop {} {\bibfield
  {journal} {\bibinfo  {journal} {Nat. Commun.}\ }\textbf {\bibinfo {volume}
  {4}},\ \bibinfo {pages} {2533} (\bibinfo {year} {2013})}\BibitemShut
  {NoStop}%
\bibitem [{\citenamefont {Xu}\ \emph {et~al.}(2016)\citenamefont {Xu},
  \citenamefont {Fegadolli}, \citenamefont {Gan}, \citenamefont {Lu},
  \citenamefont {Liu}, \citenamefont {Li}, \citenamefont {Scherer},\ and\
  \citenamefont {Chen}}]{Xu2016}%
  \BibitemOpen
  \bibfield  {author} {\bibinfo {author} {\bibfnamefont {Y.-L.}\ \bibnamefont
  {Xu}}, \bibinfo {author} {\bibfnamefont {W.~S.}\ \bibnamefont {Fegadolli}},
  \bibinfo {author} {\bibfnamefont {L.}~\bibnamefont {Gan}}, \bibinfo {author}
  {\bibfnamefont {M.-H.}\ \bibnamefont {Lu}}, \bibinfo {author} {\bibfnamefont
  {X.-P.}\ \bibnamefont {Liu}}, \bibinfo {author} {\bibfnamefont {Z.-Y.}\
  \bibnamefont {Li}}, \bibinfo {author} {\bibfnamefont {A.}~\bibnamefont
  {Scherer}}, \ and\ \bibinfo {author} {\bibfnamefont {Y.-F.}\ \bibnamefont
  {Chen}},\ }\href@noop {} {\bibfield  {journal} {\bibinfo  {journal} {Nat.
  Commun.}\ }\textbf {\bibinfo {volume} {7}},\ \bibinfo {pages} {11319}
  (\bibinfo {year} {2016})}\BibitemShut {NoStop}%
\bibitem [{\citenamefont {Xiao}\ \emph {et~al.}(2017)\citenamefont {Xiao},
  \citenamefont {Zhan}, \citenamefont {Bian}, \citenamefont {Wang},
  \citenamefont {Zhang}, \citenamefont {Wang}, \citenamefont {Li},
  \citenamefont {Mochizuki}, \citenamefont {Kim}, \citenamefont {Kawakami},
  \citenamefont {Yi}, \citenamefont {Obuse}, \citenamefont {Sandres},\ and\
  \citenamefont {Xue}}]{Xiao2017}%
  \BibitemOpen
  \bibfield  {author} {\bibinfo {author} {\bibfnamefont {L.}~\bibnamefont
  {Xiao}}, \bibinfo {author} {\bibfnamefont {X.}~\bibnamefont {Zhan}}, \bibinfo
  {author} {\bibfnamefont {Z.}~\bibnamefont {Bian}}, \bibinfo {author}
  {\bibfnamefont {K.}~\bibnamefont {Wang}}, \bibinfo {author} {\bibfnamefont
  {X.}~\bibnamefont {Zhang}}, \bibinfo {author} {\bibfnamefont
  {X.}~\bibnamefont {Wang}}, \bibinfo {author} {\bibfnamefont {J.}~\bibnamefont
  {Li}}, \bibinfo {author} {\bibfnamefont {K.}~\bibnamefont {Mochizuki}},
  \bibinfo {author} {\bibfnamefont {D.}~\bibnamefont {Kim}}, \bibinfo {author}
  {\bibfnamefont {N.}~\bibnamefont {Kawakami}}, \bibinfo {author}
  {\bibfnamefont {W.}~\bibnamefont {Yi}}, \bibinfo {author} {\bibfnamefont
  {H.}~\bibnamefont {Obuse}}, \bibinfo {author} {\bibfnamefont
  {B.}~\bibnamefont {Sandres}}, \ and\ \bibinfo {author} {\bibfnamefont
  {P.}~\bibnamefont {Xue}},\ }\href@noop {} {\bibfield  {journal} {\bibinfo
  {journal} {Nat. Phys.}\ }\textbf {\bibinfo {volume} {13}},\ \bibinfo {pages}
  {1117} (\bibinfo {year} {2017})}\BibitemShut {NoStop}%
\bibitem [{\citenamefont {Bahari}\ \emph {et~al.}(2017)\citenamefont {Bahari},
  \citenamefont {Ndao}, \citenamefont {Vallini}, \citenamefont {El~Amili},
  \citenamefont {Fainman},\ and\ \citenamefont {Kant{\'e}}}]{Bahari2017}%
  \BibitemOpen
  \bibfield  {author} {\bibinfo {author} {\bibfnamefont {B.}~\bibnamefont
  {Bahari}}, \bibinfo {author} {\bibfnamefont {A.}~\bibnamefont {Ndao}},
  \bibinfo {author} {\bibfnamefont {F.}~\bibnamefont {Vallini}}, \bibinfo
  {author} {\bibfnamefont {A.}~\bibnamefont {El~Amili}}, \bibinfo {author}
  {\bibfnamefont {Y.}~\bibnamefont {Fainman}}, \ and\ \bibinfo {author}
  {\bibfnamefont {B.}~\bibnamefont {Kant{\'e}}},\ }\href {\doibase
  10.1126/science.aao4551} {\bibfield  {journal} {\bibinfo  {journal}
  {Science}\ }\textbf {\bibinfo {volume} {358}},\ \bibinfo {pages} {636}
  (\bibinfo {year} {2017})}\BibitemShut {NoStop}%
\bibitem [{\citenamefont {Bandres}\ \emph {et~al.}(2018)\citenamefont
  {Bandres}, \citenamefont {Wittek}, \citenamefont {Harari}, \citenamefont
  {Parto}, \citenamefont {Ren}, \citenamefont {Segev}, \citenamefont
  {Christodoulides},\ and\ \citenamefont {Khajavikhan}}]{Bandres2018}%
  \BibitemOpen
  \bibfield  {author} {\bibinfo {author} {\bibfnamefont {M.~A.}\ \bibnamefont
  {Bandres}}, \bibinfo {author} {\bibfnamefont {S.}~\bibnamefont {Wittek}},
  \bibinfo {author} {\bibfnamefont {G.}~\bibnamefont {Harari}}, \bibinfo
  {author} {\bibfnamefont {M.}~\bibnamefont {Parto}}, \bibinfo {author}
  {\bibfnamefont {J.}~\bibnamefont {Ren}}, \bibinfo {author} {\bibfnamefont
  {M.}~\bibnamefont {Segev}}, \bibinfo {author} {\bibfnamefont {D.~N.}\
  \bibnamefont {Christodoulides}}, \ and\ \bibinfo {author} {\bibfnamefont
  {M.}~\bibnamefont {Khajavikhan}},\ }\href {\doibase 10.1126/science.aar4005}
  {\bibfield  {journal} {\bibinfo  {journal} {Science}\ }\textbf {\bibinfo
  {volume} {359}},\ \bibinfo {pages} {eaar4005} (\bibinfo {year}
  {2018})}\BibitemShut {NoStop}%
\bibitem [{\citenamefont {Rosenthal}\ \emph {et~al.}(2018)\citenamefont
  {Rosenthal}, \citenamefont {Ehrlich}, \citenamefont {Rudner}, \citenamefont
  {Higginbotham},\ and\ \citenamefont {Lehnert}}]{Rosenthal2018}%
  \BibitemOpen
  \bibfield  {author} {\bibinfo {author} {\bibfnamefont {E.~I.}\ \bibnamefont
  {Rosenthal}}, \bibinfo {author} {\bibfnamefont {N.~K.}\ \bibnamefont
  {Ehrlich}}, \bibinfo {author} {\bibfnamefont {M.~S.}\ \bibnamefont {Rudner}},
  \bibinfo {author} {\bibfnamefont {A.~P.}\ \bibnamefont {Higginbotham}}, \
  and\ \bibinfo {author} {\bibfnamefont {K.~W.}\ \bibnamefont {Lehnert}},\
  }\href {\doibase 10.1103/PhysRevB.97.220301} {\bibfield  {journal} {\bibinfo
  {journal} {Phys. Rev. B}\ }\textbf {\bibinfo {volume} {97}},\ \bibinfo
  {pages} {220301(R)} (\bibinfo {year} {2018})}\BibitemShut {NoStop}%
\bibitem [{\citenamefont {Li}\ \emph {et~al.}(2019{\natexlab{a}})\citenamefont
  {Li}, \citenamefont {Harter}, \citenamefont {Liu}, \citenamefont {de~Melo},
  \citenamefont {Joglekar},\ and\ \citenamefont {Luo}}]{Li2019}%
  \BibitemOpen
  \bibfield  {author} {\bibinfo {author} {\bibfnamefont {J.}~\bibnamefont
  {Li}}, \bibinfo {author} {\bibfnamefont {A.~K.}\ \bibnamefont {Harter}},
  \bibinfo {author} {\bibfnamefont {J.}~\bibnamefont {Liu}}, \bibinfo {author}
  {\bibfnamefont {L.}~\bibnamefont {de~Melo}}, \bibinfo {author} {\bibfnamefont
  {Y.~N.}\ \bibnamefont {Joglekar}}, \ and\ \bibinfo {author} {\bibfnamefont
  {L.}~\bibnamefont {Luo}},\ }\href@noop {} {\bibfield  {journal} {\bibinfo
  {journal} {Nat. Commun.}\ }\textbf {\bibinfo {volume} {10}},\ \bibinfo
  {pages} {855} (\bibinfo {year} {2019}{\natexlab{a}})}\BibitemShut {NoStop}%
\bibitem [{\citenamefont {Kremer}\ \emph {et~al.}(2019)\citenamefont {Kremer},
  \citenamefont {Biesenthal}, \citenamefont {Maczewsky}, \citenamefont
  {Heinrich}, \citenamefont {Thomale},\ and\ \citenamefont
  {Szameit}}]{Kremer2019}%
  \BibitemOpen
  \bibfield  {author} {\bibinfo {author} {\bibfnamefont {M.}~\bibnamefont
  {Kremer}}, \bibinfo {author} {\bibfnamefont {T.}~\bibnamefont {Biesenthal}},
  \bibinfo {author} {\bibfnamefont {L.~J.}\ \bibnamefont {Maczewsky}}, \bibinfo
  {author} {\bibfnamefont {M.}~\bibnamefont {Heinrich}}, \bibinfo {author}
  {\bibfnamefont {R.}~\bibnamefont {Thomale}}, \ and\ \bibinfo {author}
  {\bibfnamefont {A.}~\bibnamefont {Szameit}},\ }\href@noop {} {\bibfield
  {journal} {\bibinfo  {journal} {Nat. Commun.}\ }\textbf {\bibinfo {volume}
  {10}},\ \bibinfo {pages} {435} (\bibinfo {year} {2019})}\BibitemShut
  {NoStop}%
\bibitem [{\citenamefont {Wu}\ \emph {et~al.}(2019{\natexlab{a}})\citenamefont
  {Wu}, \citenamefont {Liu}, \citenamefont {Geng}, \citenamefont {Song},
  \citenamefont {Ye}, \citenamefont {Duan}, \citenamefont {Rong},\ and\
  \citenamefont {Du}}]{Wu2019}%
  \BibitemOpen
  \bibfield  {author} {\bibinfo {author} {\bibfnamefont {Y.}~\bibnamefont
  {Wu}}, \bibinfo {author} {\bibfnamefont {W.}~\bibnamefont {Liu}}, \bibinfo
  {author} {\bibfnamefont {J.}~\bibnamefont {Geng}}, \bibinfo {author}
  {\bibfnamefont {X.}~\bibnamefont {Song}}, \bibinfo {author} {\bibfnamefont
  {X.}~\bibnamefont {Ye}}, \bibinfo {author} {\bibfnamefont {C.-K.}\
  \bibnamefont {Duan}}, \bibinfo {author} {\bibfnamefont {X.}~\bibnamefont
  {Rong}}, \ and\ \bibinfo {author} {\bibfnamefont {J.}~\bibnamefont {Du}},\
  }\href@noop {} {\bibfield  {journal} {\bibinfo  {journal} {Science}\ }\textbf
  {\bibinfo {volume} {364}},\ \bibinfo {pages} {878} (\bibinfo {year}
  {2019}{\natexlab{a}})}\BibitemShut {NoStop}%
\bibitem [{\citenamefont {Brandenbourger}\ \emph {et~al.}(2019)\citenamefont
  {Brandenbourger}, \citenamefont {Locsin}, \citenamefont {Lerner},\ and\
  \citenamefont {Coulais}}]{Brandenbourger2019}%
  \BibitemOpen
  \bibfield  {author} {\bibinfo {author} {\bibfnamefont {M.}~\bibnamefont
  {Brandenbourger}}, \bibinfo {author} {\bibfnamefont {X.}~\bibnamefont
  {Locsin}}, \bibinfo {author} {\bibfnamefont {E.}~\bibnamefont {Lerner}}, \
  and\ \bibinfo {author} {\bibfnamefont {C.}~\bibnamefont {Coulais}},\
  }\href@noop {} {\bibfield  {journal} {\bibinfo  {journal} {Nat. Commun.}\
  }\textbf {\bibinfo {volume} {10}},\ \bibinfo {pages} {4608} (\bibinfo {year}
  {2019})}\BibitemShut {NoStop}%
\bibitem [{\citenamefont {Sakhdari}\ \emph {et~al.}(2019)\citenamefont
  {Sakhdari}, \citenamefont {Hajizadegan}, \citenamefont {Zhong}, \citenamefont
  {Christodoulides}, \citenamefont {El-Ganainy},\ and\ \citenamefont
  {Chen}}]{Sakhdari2019}%
  \BibitemOpen
  \bibfield  {author} {\bibinfo {author} {\bibfnamefont {M.}~\bibnamefont
  {Sakhdari}}, \bibinfo {author} {\bibfnamefont {M.}~\bibnamefont
  {Hajizadegan}}, \bibinfo {author} {\bibfnamefont {Q.}~\bibnamefont {Zhong}},
  \bibinfo {author} {\bibfnamefont {D.~N.}\ \bibnamefont {Christodoulides}},
  \bibinfo {author} {\bibfnamefont {R.}~\bibnamefont {El-Ganainy}}, \ and\
  \bibinfo {author} {\bibfnamefont {P.-Y.}\ \bibnamefont {Chen}},\ }\href
  {\doibase 10.1103/PhysRevLett.123.193901} {\bibfield  {journal} {\bibinfo
  {journal} {Phys. Rev. Lett.}\ }\textbf {\bibinfo {volume} {123}},\ \bibinfo
  {pages} {193901} (\bibinfo {year} {2019})}\BibitemShut {NoStop}%
\bibitem [{\citenamefont {Tuniz}\ \emph {et~al.}(2019)\citenamefont {Tuniz},
  \citenamefont {Wieduwilt},\ and\ \citenamefont {Schmidt}}]{Tuniz2019}%
  \BibitemOpen
  \bibfield  {author} {\bibinfo {author} {\bibfnamefont {A.}~\bibnamefont
  {Tuniz}}, \bibinfo {author} {\bibfnamefont {T.}~\bibnamefont {Wieduwilt}}, \
  and\ \bibinfo {author} {\bibfnamefont {M.~A.}\ \bibnamefont {Schmidt}},\
  }\href {\doibase 10.1103/PhysRevLett.123.213903} {\bibfield  {journal}
  {\bibinfo  {journal} {Phys. Rev. Lett.}\ }\textbf {\bibinfo {volume} {123}},\
  \bibinfo {pages} {213903} (\bibinfo {year} {2019})}\BibitemShut {NoStop}%
\bibitem [{\citenamefont {Xiao}\ \emph
  {et~al.}(2019{\natexlab{a}})\citenamefont {Xiao}, \citenamefont {Wang},
  \citenamefont {Zhan}, \citenamefont {Bian}, \citenamefont {Kawabata},
  \citenamefont {Ueda}, \citenamefont {Yi},\ and\ \citenamefont
  {Xue}}]{LXiao2019}%
  \BibitemOpen
  \bibfield  {author} {\bibinfo {author} {\bibfnamefont {L.}~\bibnamefont
  {Xiao}}, \bibinfo {author} {\bibfnamefont {K.}~\bibnamefont {Wang}}, \bibinfo
  {author} {\bibfnamefont {X.}~\bibnamefont {Zhan}}, \bibinfo {author}
  {\bibfnamefont {Z.}~\bibnamefont {Bian}}, \bibinfo {author} {\bibfnamefont
  {K.}~\bibnamefont {Kawabata}}, \bibinfo {author} {\bibfnamefont
  {M.}~\bibnamefont {Ueda}}, \bibinfo {author} {\bibfnamefont {W.}~\bibnamefont
  {Yi}}, \ and\ \bibinfo {author} {\bibfnamefont {P.}~\bibnamefont {Xue}},\
  }\href {\doibase 10.1103/PhysRevLett.123.230401} {\bibfield  {journal}
  {\bibinfo  {journal} {Phys. Rev. Lett.}\ }\textbf {\bibinfo {volume} {123}},\
  \bibinfo {pages} {230401} (\bibinfo {year} {2019}{\natexlab{a}})}\BibitemShut
  {NoStop}%
\bibitem [{\citenamefont {Xiao}\ \emph {et~al.}(2020)\citenamefont {Xiao},
  \citenamefont {Deng}, \citenamefont {Wang}, \citenamefont {Zhu},
  \citenamefont {Wang}, \citenamefont {Yi},\ and\ \citenamefont
  {Xue}}]{Xiao2020}%
  \BibitemOpen
  \bibfield  {author} {\bibinfo {author} {\bibfnamefont {L.}~\bibnamefont
  {Xiao}}, \bibinfo {author} {\bibfnamefont {T.}~\bibnamefont {Deng}}, \bibinfo
  {author} {\bibfnamefont {K.}~\bibnamefont {Wang}}, \bibinfo {author}
  {\bibfnamefont {G.}~\bibnamefont {Zhu}}, \bibinfo {author} {\bibfnamefont
  {Z.}~\bibnamefont {Wang}}, \bibinfo {author} {\bibfnamefont {W.}~\bibnamefont
  {Yi}}, \ and\ \bibinfo {author} {\bibfnamefont {P.}~\bibnamefont {Xue}},\
  }\href@noop {} {\bibfield  {journal} {\bibinfo  {journal} {Nat. Phys.}\
  }\textbf {\bibinfo {volume} {16}},\ \bibinfo {pages} {761} (\bibinfo {year}
  {2020})}\BibitemShut {NoStop}%
\bibitem [{\citenamefont {Poli}\ \emph {et~al.}(2015)\citenamefont {Poli},
  \citenamefont {Bellec}, \citenamefont {Kuhl}, \citenamefont {Mortessagne},\
  and\ \citenamefont {Schomerus}}]{Poli2015}%
  \BibitemOpen
  \bibfield  {author} {\bibinfo {author} {\bibfnamefont {C.}~\bibnamefont
  {Poli}}, \bibinfo {author} {\bibfnamefont {M.}~\bibnamefont {Bellec}},
  \bibinfo {author} {\bibfnamefont {U.}~\bibnamefont {Kuhl}}, \bibinfo {author}
  {\bibfnamefont {F.}~\bibnamefont {Mortessagne}}, \ and\ \bibinfo {author}
  {\bibfnamefont {H.}~\bibnamefont {Schomerus}},\ }\href@noop {} {\bibfield
  {journal} {\bibinfo  {journal} {Nat. Commun.}\ }\textbf {\bibinfo {volume}
  {6}},\ \bibinfo {pages} {6710} (\bibinfo {year} {2015})}\BibitemShut
  {NoStop}%
\bibitem [{\citenamefont {Zeuner}\ \emph {et~al.}(2015)\citenamefont {Zeuner},
  \citenamefont {Rechtsman}, \citenamefont {Plotnik}, \citenamefont {Lumer},
  \citenamefont {Nolte}, \citenamefont {Rudner}, \citenamefont {Segev},\ and\
  \citenamefont {Szameit}}]{Zeuner2015}%
  \BibitemOpen
  \bibfield  {author} {\bibinfo {author} {\bibfnamefont {J.~M.}\ \bibnamefont
  {Zeuner}}, \bibinfo {author} {\bibfnamefont {M.~C.}\ \bibnamefont
  {Rechtsman}}, \bibinfo {author} {\bibfnamefont {Y.}~\bibnamefont {Plotnik}},
  \bibinfo {author} {\bibfnamefont {Y.}~\bibnamefont {Lumer}}, \bibinfo
  {author} {\bibfnamefont {S.}~\bibnamefont {Nolte}}, \bibinfo {author}
  {\bibfnamefont {M.~S.}\ \bibnamefont {Rudner}}, \bibinfo {author}
  {\bibfnamefont {M.}~\bibnamefont {Segev}}, \ and\ \bibinfo {author}
  {\bibfnamefont {A.}~\bibnamefont {Szameit}},\ }\href {\doibase
  10.1103/PhysRevLett.115.040402} {\bibfield  {journal} {\bibinfo  {journal}
  {Phys. Rev. Lett.}\ }\textbf {\bibinfo {volume} {115}},\ \bibinfo {pages}
  {040402} (\bibinfo {year} {2015})}\BibitemShut {NoStop}%
\bibitem [{\citenamefont {Weimann}\ \emph {et~al.}(2017)\citenamefont
  {Weimann}, \citenamefont {Kremer}, \citenamefont {Plotnik}, \citenamefont
  {Lumer}, \citenamefont {Nolte}, \citenamefont {Makris}, \citenamefont
  {Segev}, \citenamefont {Rechtsman},\ and\ \citenamefont
  {Szameit}}]{Weimann2017}%
  \BibitemOpen
  \bibfield  {author} {\bibinfo {author} {\bibfnamefont {S.}~\bibnamefont
  {Weimann}}, \bibinfo {author} {\bibfnamefont {M.}~\bibnamefont {Kremer}},
  \bibinfo {author} {\bibfnamefont {Y.}~\bibnamefont {Plotnik}}, \bibinfo
  {author} {\bibfnamefont {Y.}~\bibnamefont {Lumer}}, \bibinfo {author}
  {\bibfnamefont {S.}~\bibnamefont {Nolte}}, \bibinfo {author} {\bibfnamefont
  {K.}~\bibnamefont {Makris}}, \bibinfo {author} {\bibfnamefont
  {M.}~\bibnamefont {Segev}}, \bibinfo {author} {\bibfnamefont
  {M.}~\bibnamefont {Rechtsman}}, \ and\ \bibinfo {author} {\bibfnamefont
  {A.}~\bibnamefont {Szameit}},\ }\href@noop {} {\bibfield  {journal} {\bibinfo
   {journal} {Nat. Mater.}\ }\textbf {\bibinfo {volume} {16}},\ \bibinfo
  {pages} {433} (\bibinfo {year} {2017})}\BibitemShut {NoStop}%
\bibitem [{\citenamefont {St-Jean}\ \emph {et~al.}(2017)\citenamefont
  {St-Jean}, \citenamefont {Goblot}, \citenamefont {Galopin}, \citenamefont
  {Lema{\^\i}tre}, \citenamefont {Ozawa}, \citenamefont {Le~Gratiet},
  \citenamefont {Sagnes}, \citenamefont {Bloch},\ and\ \citenamefont
  {Amo}}]{St2017}%
  \BibitemOpen
  \bibfield  {author} {\bibinfo {author} {\bibfnamefont {P.}~\bibnamefont
  {St-Jean}}, \bibinfo {author} {\bibfnamefont {V.}~\bibnamefont {Goblot}},
  \bibinfo {author} {\bibfnamefont {E.}~\bibnamefont {Galopin}}, \bibinfo
  {author} {\bibfnamefont {A.}~\bibnamefont {Lema{\^\i}tre}}, \bibinfo {author}
  {\bibfnamefont {T.}~\bibnamefont {Ozawa}}, \bibinfo {author} {\bibfnamefont
  {L.}~\bibnamefont {Le~Gratiet}}, \bibinfo {author} {\bibfnamefont
  {I.}~\bibnamefont {Sagnes}}, \bibinfo {author} {\bibfnamefont
  {J.}~\bibnamefont {Bloch}}, \ and\ \bibinfo {author} {\bibfnamefont
  {A.}~\bibnamefont {Amo}},\ }\href@noop {} {\bibfield  {journal} {\bibinfo
  {journal} {Nat. Photonics}\ }\textbf {\bibinfo {volume} {11}},\ \bibinfo
  {pages} {651} (\bibinfo {year} {2017})}\BibitemShut {NoStop}%
\bibitem [{\citenamefont {Parto}\ \emph {et~al.}(2018)\citenamefont {Parto},
  \citenamefont {Wittek}, \citenamefont {Hodaei}, \citenamefont {Harari},
  \citenamefont {Bandres}, \citenamefont {Ren}, \citenamefont {Rechtsman},
  \citenamefont {Segev}, \citenamefont {Christodoulides},\ and\ \citenamefont
  {Khajavikhan}}]{Parto2018}%
  \BibitemOpen
  \bibfield  {author} {\bibinfo {author} {\bibfnamefont {M.}~\bibnamefont
  {Parto}}, \bibinfo {author} {\bibfnamefont {S.}~\bibnamefont {Wittek}},
  \bibinfo {author} {\bibfnamefont {H.}~\bibnamefont {Hodaei}}, \bibinfo
  {author} {\bibfnamefont {G.}~\bibnamefont {Harari}}, \bibinfo {author}
  {\bibfnamefont {M.~A.}\ \bibnamefont {Bandres}}, \bibinfo {author}
  {\bibfnamefont {J.}~\bibnamefont {Ren}}, \bibinfo {author} {\bibfnamefont
  {M.~C.}\ \bibnamefont {Rechtsman}}, \bibinfo {author} {\bibfnamefont
  {M.}~\bibnamefont {Segev}}, \bibinfo {author} {\bibfnamefont {D.~N.}\
  \bibnamefont {Christodoulides}}, \ and\ \bibinfo {author} {\bibfnamefont
  {M.}~\bibnamefont {Khajavikhan}},\ }\href {\doibase
  10.1103/PhysRevLett.120.113901} {\bibfield  {journal} {\bibinfo  {journal}
  {Phys. Rev. Lett.}\ }\textbf {\bibinfo {volume} {120}},\ \bibinfo {pages}
  {113901} (\bibinfo {year} {2018})}\BibitemShut {NoStop}%
\bibitem [{\citenamefont {Pan}\ \emph {et~al.}(2018)\citenamefont {Pan},
  \citenamefont {Zhao}, \citenamefont {Miao}, \citenamefont {Longhi},\ and\
  \citenamefont {Feng}}]{Pan2018}%
  \BibitemOpen
  \bibfield  {author} {\bibinfo {author} {\bibfnamefont {M.}~\bibnamefont
  {Pan}}, \bibinfo {author} {\bibfnamefont {H.}~\bibnamefont {Zhao}}, \bibinfo
  {author} {\bibfnamefont {P.}~\bibnamefont {Miao}}, \bibinfo {author}
  {\bibfnamefont {S.}~\bibnamefont {Longhi}}, \ and\ \bibinfo {author}
  {\bibfnamefont {L.}~\bibnamefont {Feng}},\ }\href@noop {} {\bibfield
  {journal} {\bibinfo  {journal} {Nat. Commun.}\ }\textbf {\bibinfo {volume}
  {9}},\ \bibinfo {pages} {1308} (\bibinfo {year} {2018})}\BibitemShut
  {NoStop}%
\bibitem [{\citenamefont {Ghatak}\ \emph {et~al.}()\citenamefont {Ghatak},
  \citenamefont {Brandenbourger}, \citenamefont {van Wezel},\ and\
  \citenamefont {Coulais}}]{Ghatak2019e}%
  \BibitemOpen
  \bibfield  {author} {\bibinfo {author} {\bibfnamefont {A.}~\bibnamefont
  {Ghatak}}, \bibinfo {author} {\bibfnamefont {M.}~\bibnamefont
  {Brandenbourger}}, \bibinfo {author} {\bibfnamefont {J.}~\bibnamefont {van
  Wezel}}, \ and\ \bibinfo {author} {\bibfnamefont {C.}~\bibnamefont
  {Coulais}},\ }\href@noop {} {\bibinfo  {journal} {arXiv:1907.11619}\
  }\BibitemShut {NoStop}%
\bibitem [{\citenamefont {Helbig}\ \emph {et~al.}(2020)\citenamefont {Helbig},
  \citenamefont {Hofmann}, \citenamefont {Imhof}, \citenamefont {Abdelghany},
  \citenamefont {Kiessling}, \citenamefont {Molenkamp}, \citenamefont {Lee},
  \citenamefont {Szameit}, \citenamefont {Greiter},\ and\ \citenamefont
  {Thomale}}]{Helbig2020}%
  \BibitemOpen
\bibfield  {journal} {  }\bibfield  {author} {\bibinfo {author} {\bibfnamefont
  {T.}~\bibnamefont {Helbig}}, \bibinfo {author} {\bibfnamefont
  {T.}~\bibnamefont {Hofmann}}, \bibinfo {author} {\bibfnamefont
  {S.}~\bibnamefont {Imhof}}, \bibinfo {author} {\bibfnamefont
  {M.}~\bibnamefont {Abdelghany}}, \bibinfo {author} {\bibfnamefont
  {T.}~\bibnamefont {Kiessling}}, \bibinfo {author} {\bibfnamefont
  {L.}~\bibnamefont {Molenkamp}}, \bibinfo {author} {\bibfnamefont
  {C.}~\bibnamefont {Lee}}, \bibinfo {author} {\bibfnamefont {A.}~\bibnamefont
  {Szameit}}, \bibinfo {author} {\bibfnamefont {M.}~\bibnamefont {Greiter}}, \
  and\ \bibinfo {author} {\bibfnamefont {R.}~\bibnamefont {Thomale}},\
  }\href@noop {} {\bibfield  {journal} {\bibinfo  {journal} {Nat. Phys.}\
  }\textbf {\bibinfo {volume} {16}},\ \bibinfo {pages} {747} (\bibinfo {year}
  {2020})}\BibitemShut {NoStop}%
\bibitem [{\citenamefont {Berry}(2004)}]{Berry2004}%
  \BibitemOpen
  \bibfield  {author} {\bibinfo {author} {\bibfnamefont {M.~V.}\ \bibnamefont
  {Berry}},\ }\href@noop {} {\bibfield  {journal} {\bibinfo  {journal} {Czech.
  J. Phys.}\ }\textbf {\bibinfo {volume} {54}},\ \bibinfo {pages} {1039}
  (\bibinfo {year} {2004})}\BibitemShut {NoStop}%
\bibitem [{\citenamefont {Heiss}(2012)}]{Heiss2012}%
  \BibitemOpen
  \bibfield  {author} {\bibinfo {author} {\bibfnamefont {W.}~\bibnamefont
  {Heiss}},\ }\href@noop {} {\bibfield  {journal} {\bibinfo  {journal} {J.
  Phys. A}\ }\textbf {\bibinfo {volume} {45}},\ \bibinfo {pages} {444016}
  (\bibinfo {year} {2012})}\BibitemShut {NoStop}%
\bibitem [{\citenamefont {Dembowski}\ \emph {et~al.}(2001)\citenamefont
  {Dembowski}, \citenamefont {Gr\"af}, \citenamefont {Harney}, \citenamefont
  {Heine}, \citenamefont {Heiss}, \citenamefont {Rehfeld},\ and\ \citenamefont
  {Richter}}]{Dembowski2001}%
  \BibitemOpen
  \bibfield  {author} {\bibinfo {author} {\bibfnamefont {C.}~\bibnamefont
  {Dembowski}}, \bibinfo {author} {\bibfnamefont {H.-D.}\ \bibnamefont
  {Gr\"af}}, \bibinfo {author} {\bibfnamefont {H.~L.}\ \bibnamefont {Harney}},
  \bibinfo {author} {\bibfnamefont {A.}~\bibnamefont {Heine}}, \bibinfo
  {author} {\bibfnamefont {W.~D.}\ \bibnamefont {Heiss}}, \bibinfo {author}
  {\bibfnamefont {H.}~\bibnamefont {Rehfeld}}, \ and\ \bibinfo {author}
  {\bibfnamefont {A.}~\bibnamefont {Richter}},\ }\href {\doibase
  10.1103/PhysRevLett.86.787} {\bibfield  {journal} {\bibinfo  {journal} {Phys.
  Rev. Lett.}\ }\textbf {\bibinfo {volume} {86}},\ \bibinfo {pages} {787}
  (\bibinfo {year} {2001})}\BibitemShut {NoStop}%
\bibitem [{\citenamefont {Zhen}\ \emph {et~al.}(2015)\citenamefont {Zhen},
  \citenamefont {Hsu}, \citenamefont {Igarashi}, \citenamefont {Lu},
  \citenamefont {Kaminer}, \citenamefont {Pick}, \citenamefont {Chua},
  \citenamefont {Joannopoulos},\ and\ \citenamefont
  {Solja{\v{c}}i{\'c}}}]{Zhen2015}%
  \BibitemOpen
  \bibfield  {author} {\bibinfo {author} {\bibfnamefont {B.}~\bibnamefont
  {Zhen}}, \bibinfo {author} {\bibfnamefont {C.~W.}\ \bibnamefont {Hsu}},
  \bibinfo {author} {\bibfnamefont {Y.}~\bibnamefont {Igarashi}}, \bibinfo
  {author} {\bibfnamefont {L.}~\bibnamefont {Lu}}, \bibinfo {author}
  {\bibfnamefont {I.}~\bibnamefont {Kaminer}}, \bibinfo {author} {\bibfnamefont
  {A.}~\bibnamefont {Pick}}, \bibinfo {author} {\bibfnamefont {S.-L.}\
  \bibnamefont {Chua}}, \bibinfo {author} {\bibfnamefont {J.~D.}\ \bibnamefont
  {Joannopoulos}}, \ and\ \bibinfo {author} {\bibfnamefont {M.}~\bibnamefont
  {Solja{\v{c}}i{\'c}}},\ }\href@noop {} {\bibfield  {journal} {\bibinfo
  {journal} {Nature (London)}\ }\textbf {\bibinfo {volume} {525}},\ \bibinfo
  {pages} {354} (\bibinfo {year} {2015})}\BibitemShut {NoStop}%
\bibitem [{\citenamefont {Ding}\ \emph {et~al.}(2016)\citenamefont {Ding},
  \citenamefont {Ma}, \citenamefont {Xiao}, \citenamefont {Zhang},\ and\
  \citenamefont {Chan}}]{Ding2016}%
  \BibitemOpen
  \bibfield  {author} {\bibinfo {author} {\bibfnamefont {K.}~\bibnamefont
  {Ding}}, \bibinfo {author} {\bibfnamefont {G.}~\bibnamefont {Ma}}, \bibinfo
  {author} {\bibfnamefont {M.}~\bibnamefont {Xiao}}, \bibinfo {author}
  {\bibfnamefont {Z.~Q.}\ \bibnamefont {Zhang}}, \ and\ \bibinfo {author}
  {\bibfnamefont {C.~T.}\ \bibnamefont {Chan}},\ }\href {\doibase
  10.1103/PhysRevX.6.021007} {\bibfield  {journal} {\bibinfo  {journal} {Phys.
  Rev. X}\ }\textbf {\bibinfo {volume} {6}},\ \bibinfo {pages} {021007}
  (\bibinfo {year} {2016})}\BibitemShut {NoStop}%
\bibitem [{\citenamefont {Zhou}\ \emph {et~al.}(2018)\citenamefont {Zhou},
  \citenamefont {Peng}, \citenamefont {Yoon}, \citenamefont {Hsu},
  \citenamefont {Nelson}, \citenamefont {Fu}, \citenamefont {Joannopoulos},
  \citenamefont {Solja{\v c}i{\'c}},\ and\ \citenamefont {Zhen}}]{Zhou2018}%
  \BibitemOpen
  \bibfield  {author} {\bibinfo {author} {\bibfnamefont {H.}~\bibnamefont
  {Zhou}}, \bibinfo {author} {\bibfnamefont {C.}~\bibnamefont {Peng}}, \bibinfo
  {author} {\bibfnamefont {Y.}~\bibnamefont {Yoon}}, \bibinfo {author}
  {\bibfnamefont {C.~W.}\ \bibnamefont {Hsu}}, \bibinfo {author} {\bibfnamefont
  {K.~A.}\ \bibnamefont {Nelson}}, \bibinfo {author} {\bibfnamefont
  {L.}~\bibnamefont {Fu}}, \bibinfo {author} {\bibfnamefont {J.~D.}\
  \bibnamefont {Joannopoulos}}, \bibinfo {author} {\bibfnamefont
  {M.}~\bibnamefont {Solja{\v c}i{\'c}}}, \ and\ \bibinfo {author}
  {\bibfnamefont {B.}~\bibnamefont {Zhen}},\ }\href {\doibase
  10.1126/science.aap9859} {\bibfield  {journal} {\bibinfo  {journal}
  {Science}\ }\textbf {\bibinfo {volume} {359}},\ \bibinfo {pages} {1009}
  (\bibinfo {year} {2018})}\BibitemShut {NoStop}%
\bibitem [{\citenamefont {Ding}\ \emph {et~al.}(2018)\citenamefont {Ding},
  \citenamefont {Ma}, \citenamefont {Zhang},\ and\ \citenamefont
  {Chan}}]{Ding2018}%
  \BibitemOpen
  \bibfield  {author} {\bibinfo {author} {\bibfnamefont {K.}~\bibnamefont
  {Ding}}, \bibinfo {author} {\bibfnamefont {G.}~\bibnamefont {Ma}}, \bibinfo
  {author} {\bibfnamefont {Z.~Q.}\ \bibnamefont {Zhang}}, \ and\ \bibinfo
  {author} {\bibfnamefont {C.~T.}\ \bibnamefont {Chan}},\ }\href {\doibase
  10.1103/PhysRevLett.121.085702} {\bibfield  {journal} {\bibinfo  {journal}
  {Phys. Rev. Lett.}\ }\textbf {\bibinfo {volume} {121}},\ \bibinfo {pages}
  {085702} (\bibinfo {year} {2018})}\BibitemShut {NoStop}%
\bibitem [{\citenamefont {Cerjan}\ \emph {et~al.}(2019)\citenamefont {Cerjan},
  \citenamefont {Huang}, \citenamefont {Wang}, \citenamefont {Chen},
  \citenamefont {Chong},\ and\ \citenamefont {Rechtsman}}]{Cerjan2019}%
  \BibitemOpen
  \bibfield  {author} {\bibinfo {author} {\bibfnamefont {A.}~\bibnamefont
  {Cerjan}}, \bibinfo {author} {\bibfnamefont {S.}~\bibnamefont {Huang}},
  \bibinfo {author} {\bibfnamefont {M.}~\bibnamefont {Wang}}, \bibinfo {author}
  {\bibfnamefont {K.~P.}\ \bibnamefont {Chen}}, \bibinfo {author}
  {\bibfnamefont {Y.}~\bibnamefont {Chong}}, \ and\ \bibinfo {author}
  {\bibfnamefont {M.~C.}\ \bibnamefont {Rechtsman}},\ }\href@noop {} {\bibfield
   {journal} {\bibinfo  {journal} {Nat. Photonics}\ }\textbf {\bibinfo {volume}
  {13}},\ \bibinfo {pages} {623} (\bibinfo {year} {2019})}\BibitemShut
  {NoStop}%
\bibitem [{\citenamefont {Zhang}\ \emph
  {et~al.}(2019{\natexlab{a}})\citenamefont {Zhang}, \citenamefont {Ding},
  \citenamefont {Zhou}, \citenamefont {Xu},\ and\ \citenamefont
  {Jin}}]{XZhang2019}%
  \BibitemOpen
  \bibfield  {author} {\bibinfo {author} {\bibfnamefont {X.}~\bibnamefont
  {Zhang}}, \bibinfo {author} {\bibfnamefont {K.}~\bibnamefont {Ding}},
  \bibinfo {author} {\bibfnamefont {X.}~\bibnamefont {Zhou}}, \bibinfo {author}
  {\bibfnamefont {J.}~\bibnamefont {Xu}}, \ and\ \bibinfo {author}
  {\bibfnamefont {D.}~\bibnamefont {Jin}},\ }\href {\doibase
  10.1103/PhysRevLett.123.237202} {\bibfield  {journal} {\bibinfo  {journal}
  {Phys. Rev. Lett.}\ }\textbf {\bibinfo {volume} {123}},\ \bibinfo {pages}
  {237202} (\bibinfo {year} {2019}{\natexlab{a}})}\BibitemShut {NoStop}%
\bibitem [{\citenamefont {Guo}\ \emph {et~al.}(2009)\citenamefont {Guo},
  \citenamefont {Salamo}, \citenamefont {Duchesne}, \citenamefont {Morandotti},
  \citenamefont {Volatier-Ravat}, \citenamefont {Aimez}, \citenamefont
  {Siviloglou},\ and\ \citenamefont {Christodoulides}}]{Guo2009}%
  \BibitemOpen
  \bibfield  {author} {\bibinfo {author} {\bibfnamefont {A.}~\bibnamefont
  {Guo}}, \bibinfo {author} {\bibfnamefont {G.~J.}\ \bibnamefont {Salamo}},
  \bibinfo {author} {\bibfnamefont {D.}~\bibnamefont {Duchesne}}, \bibinfo
  {author} {\bibfnamefont {R.}~\bibnamefont {Morandotti}}, \bibinfo {author}
  {\bibfnamefont {M.}~\bibnamefont {Volatier-Ravat}}, \bibinfo {author}
  {\bibfnamefont {V.}~\bibnamefont {Aimez}}, \bibinfo {author} {\bibfnamefont
  {G.~A.}\ \bibnamefont {Siviloglou}}, \ and\ \bibinfo {author} {\bibfnamefont
  {D.~N.}\ \bibnamefont {Christodoulides}},\ }\href {\doibase
  10.1103/PhysRevLett.103.093902} {\bibfield  {journal} {\bibinfo  {journal}
  {Phys. Rev. Lett.}\ }\textbf {\bibinfo {volume} {103}},\ \bibinfo {pages}
  {093902} (\bibinfo {year} {2009})}\BibitemShut {NoStop}%
\bibitem [{\citenamefont {Feng}\ \emph {et~al.}(2013)\citenamefont {Feng},
  \citenamefont {Xu}, \citenamefont {Fegadolli}, \citenamefont {Lu},
  \citenamefont {Oliveira}, \citenamefont {Almeida}, \citenamefont {Chen},\
  and\ \citenamefont {Scherer}}]{Feng2013}%
  \BibitemOpen
  \bibfield  {author} {\bibinfo {author} {\bibfnamefont {L.}~\bibnamefont
  {Feng}}, \bibinfo {author} {\bibfnamefont {Y.-L.}\ \bibnamefont {Xu}},
  \bibinfo {author} {\bibfnamefont {W.~S.}\ \bibnamefont {Fegadolli}}, \bibinfo
  {author} {\bibfnamefont {M.-H.}\ \bibnamefont {Lu}}, \bibinfo {author}
  {\bibfnamefont {J.~E.}\ \bibnamefont {Oliveira}}, \bibinfo {author}
  {\bibfnamefont {V.~R.}\ \bibnamefont {Almeida}}, \bibinfo {author}
  {\bibfnamefont {Y.-F.}\ \bibnamefont {Chen}}, \ and\ \bibinfo {author}
  {\bibfnamefont {A.}~\bibnamefont {Scherer}},\ }\href@noop {} {\bibfield
  {journal} {\bibinfo  {journal} {Nat. Mater.}\ }\textbf {\bibinfo {volume}
  {12}},\ \bibinfo {pages} {108} (\bibinfo {year} {2013})}\BibitemShut
  {NoStop}%
\bibitem [{\citenamefont {Brandstetter}\ \emph {et~al.}(2014)\citenamefont
  {Brandstetter}, \citenamefont {Liertzer}, \citenamefont {Deutsch},
  \citenamefont {Klang}, \citenamefont {Sch{\"o}berl}, \citenamefont
  {T{\"u}reci}, \citenamefont {Strasser}, \citenamefont {Unterrainer},\ and\
  \citenamefont {Rotter}}]{Brandstetter2014}%
  \BibitemOpen
  \bibfield  {author} {\bibinfo {author} {\bibfnamefont {M.}~\bibnamefont
  {Brandstetter}}, \bibinfo {author} {\bibfnamefont {M.}~\bibnamefont
  {Liertzer}}, \bibinfo {author} {\bibfnamefont {C.}~\bibnamefont {Deutsch}},
  \bibinfo {author} {\bibfnamefont {P.}~\bibnamefont {Klang}}, \bibinfo
  {author} {\bibfnamefont {J.}~\bibnamefont {Sch{\"o}berl}}, \bibinfo {author}
  {\bibfnamefont {H.}~\bibnamefont {T{\"u}reci}}, \bibinfo {author}
  {\bibfnamefont {G.}~\bibnamefont {Strasser}}, \bibinfo {author}
  {\bibfnamefont {K.}~\bibnamefont {Unterrainer}}, \ and\ \bibinfo {author}
  {\bibfnamefont {S.}~\bibnamefont {Rotter}},\ }\href@noop {} {\bibfield
  {journal} {\bibinfo  {journal} {Nat. Commun.}\ }\textbf {\bibinfo {volume}
  {5}},\ \bibinfo {pages} {4034} (\bibinfo {year} {2014})}\BibitemShut
  {NoStop}%
\bibitem [{\citenamefont {Chen}\ \emph {et~al.}(2017)\citenamefont {Chen},
  \citenamefont {{\"O}zdemir}, \citenamefont {Zhao}, \citenamefont {Wiersig},\
  and\ \citenamefont {Yang}}]{Chen2017}%
  \BibitemOpen
  \bibfield  {author} {\bibinfo {author} {\bibfnamefont {W.}~\bibnamefont
  {Chen}}, \bibinfo {author} {\bibfnamefont {{\c{S}}.~K.}\ \bibnamefont
  {{\"O}zdemir}}, \bibinfo {author} {\bibfnamefont {G.}~\bibnamefont {Zhao}},
  \bibinfo {author} {\bibfnamefont {J.}~\bibnamefont {Wiersig}}, \ and\
  \bibinfo {author} {\bibfnamefont {L.}~\bibnamefont {Yang}},\ }\href@noop {}
  {\bibfield  {journal} {\bibinfo  {journal} {Nature (London)}\ }\textbf
  {\bibinfo {volume} {548}},\ \bibinfo {pages} {192} (\bibinfo {year}
  {2017})}\BibitemShut {NoStop}%
\bibitem [{\citenamefont {Wang}\ \emph
  {et~al.}(2019{\natexlab{a}})\citenamefont {Wang}, \citenamefont {Fang},
  \citenamefont {Mao}, \citenamefont {Jing},\ and\ \citenamefont
  {Li}}]{XWang2019}%
  \BibitemOpen
  \bibfield  {author} {\bibinfo {author} {\bibfnamefont {X.}~\bibnamefont
  {Wang}}, \bibinfo {author} {\bibfnamefont {X.}~\bibnamefont {Fang}}, \bibinfo
  {author} {\bibfnamefont {D.}~\bibnamefont {Mao}}, \bibinfo {author}
  {\bibfnamefont {Y.}~\bibnamefont {Jing}}, \ and\ \bibinfo {author}
  {\bibfnamefont {Y.}~\bibnamefont {Li}},\ }\href {\doibase
  10.1103/PhysRevLett.123.214302} {\bibfield  {journal} {\bibinfo  {journal}
  {Phys. Rev. Lett.}\ }\textbf {\bibinfo {volume} {123}},\ \bibinfo {pages}
  {214302} (\bibinfo {year} {2019}{\natexlab{a}})}\BibitemShut {NoStop}%
\bibitem [{\citenamefont {Ding}\ \emph {et~al.}(2015)\citenamefont {Ding},
  \citenamefont {Zhang},\ and\ \citenamefont {Chan}}]{Ding2015}%
  \BibitemOpen
  \bibfield  {author} {\bibinfo {author} {\bibfnamefont {K.}~\bibnamefont
  {Ding}}, \bibinfo {author} {\bibfnamefont {Z.~Q.}\ \bibnamefont {Zhang}}, \
  and\ \bibinfo {author} {\bibfnamefont {C.~T.}\ \bibnamefont {Chan}},\ }\href
  {\doibase 10.1103/PhysRevB.92.235310} {\bibfield  {journal} {\bibinfo
  {journal} {Phys. Rev. B}\ }\textbf {\bibinfo {volume} {92}},\ \bibinfo
  {pages} {235310} (\bibinfo {year} {2015})}\BibitemShut {NoStop}%
\bibitem [{\citenamefont {Lee}(2016)}]{Lee2016}%
  \BibitemOpen
  \bibfield  {author} {\bibinfo {author} {\bibfnamefont {T.~E.}\ \bibnamefont
  {Lee}},\ }\href {\doibase 10.1103/PhysRevLett.116.133903} {\bibfield
  {journal} {\bibinfo  {journal} {Phys. Rev. Lett.}\ }\textbf {\bibinfo
  {volume} {116}},\ \bibinfo {pages} {133903} (\bibinfo {year}
  {2016})}\BibitemShut {NoStop}%
\bibitem [{\citenamefont {Lin}\ \emph {et~al.}(2016)\citenamefont {Lin},
  \citenamefont {Pick}, \citenamefont {Lon\ifmmode~\check{c}\else
  \v{c}\fi{}ar},\ and\ \citenamefont {Rodriguez}}]{Lin2016}%
  \BibitemOpen
  \bibfield  {author} {\bibinfo {author} {\bibfnamefont {Z.}~\bibnamefont
  {Lin}}, \bibinfo {author} {\bibfnamefont {A.}~\bibnamefont {Pick}}, \bibinfo
  {author} {\bibfnamefont {M.}~\bibnamefont {Lon\ifmmode~\check{c}\else
  \v{c}\fi{}ar}}, \ and\ \bibinfo {author} {\bibfnamefont {A.~W.}\ \bibnamefont
  {Rodriguez}},\ }\href {\doibase 10.1103/PhysRevLett.117.107402} {\bibfield
  {journal} {\bibinfo  {journal} {Phys. Rev. Lett.}\ }\textbf {\bibinfo
  {volume} {117}},\ \bibinfo {pages} {107402} (\bibinfo {year}
  {2016})}\BibitemShut {NoStop}%
\bibitem [{\citenamefont {Leykam}\ \emph {et~al.}(2017)\citenamefont {Leykam},
  \citenamefont {Bliokh}, \citenamefont {Huang}, \citenamefont {Chong},\ and\
  \citenamefont {Nori}}]{Leykam2017}%
  \BibitemOpen
  \bibfield  {author} {\bibinfo {author} {\bibfnamefont {D.}~\bibnamefont
  {Leykam}}, \bibinfo {author} {\bibfnamefont {K.~Y.}\ \bibnamefont {Bliokh}},
  \bibinfo {author} {\bibfnamefont {C.}~\bibnamefont {Huang}}, \bibinfo
  {author} {\bibfnamefont {Y.~D.}\ \bibnamefont {Chong}}, \ and\ \bibinfo
  {author} {\bibfnamefont {F.}~\bibnamefont {Nori}},\ }\href {\doibase
  10.1103/PhysRevLett.118.040401} {\bibfield  {journal} {\bibinfo  {journal}
  {Phys. Rev. Lett.}\ }\textbf {\bibinfo {volume} {118}},\ \bibinfo {pages}
  {040401} (\bibinfo {year} {2017})}\BibitemShut {NoStop}%
\bibitem [{\citenamefont {Xu}\ \emph {et~al.}(2017)\citenamefont {Xu},
  \citenamefont {Wang},\ and\ \citenamefont {Duan}}]{Xu2017}%
  \BibitemOpen
  \bibfield  {author} {\bibinfo {author} {\bibfnamefont {Y.}~\bibnamefont
  {Xu}}, \bibinfo {author} {\bibfnamefont {S.-T.}\ \bibnamefont {Wang}}, \ and\
  \bibinfo {author} {\bibfnamefont {L.-M.}\ \bibnamefont {Duan}},\ }\href
  {\doibase 10.1103/PhysRevLett.118.045701} {\bibfield  {journal} {\bibinfo
  {journal} {Phys. Rev. Lett.}\ }\textbf {\bibinfo {volume} {118}},\ \bibinfo
  {pages} {045701} (\bibinfo {year} {2017})}\BibitemShut {NoStop}%
\bibitem [{\citenamefont {Gonz\'alez}\ and\ \citenamefont
  {Molina}(2017)}]{Gonzalez2017}%
  \BibitemOpen
  \bibfield  {author} {\bibinfo {author} {\bibfnamefont {J.}~\bibnamefont
  {Gonz\'alez}}\ and\ \bibinfo {author} {\bibfnamefont {R.~A.}\ \bibnamefont
  {Molina}},\ }\href {\doibase 10.1103/PhysRevB.96.045437} {\bibfield
  {journal} {\bibinfo  {journal} {Phys. Rev. B}\ }\textbf {\bibinfo {volume}
  {96}},\ \bibinfo {pages} {045437} (\bibinfo {year} {2017})}\BibitemShut
  {NoStop}%
\bibitem [{\citenamefont {Zyuzin}\ and\ \citenamefont
  {Zyuzin}(2018)}]{Zyuzin2018}%
  \BibitemOpen
  \bibfield  {author} {\bibinfo {author} {\bibfnamefont {A.~A.}\ \bibnamefont
  {Zyuzin}}\ and\ \bibinfo {author} {\bibfnamefont {A.~Y.}\ \bibnamefont
  {Zyuzin}},\ }\href {\doibase 10.1103/PhysRevB.97.041203} {\bibfield
  {journal} {\bibinfo  {journal} {Phys. Rev. B}\ }\textbf {\bibinfo {volume}
  {97}},\ \bibinfo {pages} {041203(R)} (\bibinfo {year} {2018})}\BibitemShut
  {NoStop}%
\bibitem [{\citenamefont {Wang}\ \emph {et~al.}(2018)\citenamefont {Wang},
  \citenamefont {Dong}, \citenamefont {Shi}, \citenamefont {Wang},\ and\
  \citenamefont {Fung}}]{Wang2018}%
  \BibitemOpen
  \bibfield  {author} {\bibinfo {author} {\bibfnamefont {J.}~\bibnamefont
  {Wang}}, \bibinfo {author} {\bibfnamefont {H.~Y.}\ \bibnamefont {Dong}},
  \bibinfo {author} {\bibfnamefont {Q.~Y.}\ \bibnamefont {Shi}}, \bibinfo
  {author} {\bibfnamefont {W.}~\bibnamefont {Wang}}, \ and\ \bibinfo {author}
  {\bibfnamefont {K.~H.}\ \bibnamefont {Fung}},\ }\href {\doibase
  10.1103/PhysRevB.97.014428} {\bibfield  {journal} {\bibinfo  {journal} {Phys.
  Rev. B}\ }\textbf {\bibinfo {volume} {97}},\ \bibinfo {pages} {014428}
  (\bibinfo {year} {2018})}\BibitemShut {NoStop}%
\bibitem [{\citenamefont {Martinez~Alvarez}\ \emph {et~al.}(2018)\citenamefont
  {Martinez~Alvarez}, \citenamefont {Barrios~Vargas},\ and\ \citenamefont
  {Foa~Torres}}]{Martinez2018}%
  \BibitemOpen
  \bibfield  {author} {\bibinfo {author} {\bibfnamefont {V.~M.}\ \bibnamefont
  {Martinez~Alvarez}}, \bibinfo {author} {\bibfnamefont {J.~E.}\ \bibnamefont
  {Barrios~Vargas}}, \ and\ \bibinfo {author} {\bibfnamefont {L.~E.~F.}\
  \bibnamefont {Foa~Torres}},\ }\href {\doibase 10.1103/PhysRevB.97.121401}
  {\bibfield  {journal} {\bibinfo  {journal} {Phys. Rev. B}\ }\textbf {\bibinfo
  {volume} {97}},\ \bibinfo {pages} {121401(R)} (\bibinfo {year}
  {2018})}\BibitemShut {NoStop}%
\bibitem [{\citenamefont {Molina}\ and\ \citenamefont
  {Gonz\'alez}(2018)}]{Molina2018}%
  \BibitemOpen
  \bibfield  {author} {\bibinfo {author} {\bibfnamefont {R.~A.}\ \bibnamefont
  {Molina}}\ and\ \bibinfo {author} {\bibfnamefont {J.}~\bibnamefont
  {Gonz\'alez}},\ }\href {\doibase 10.1103/PhysRevLett.120.146601} {\bibfield
  {journal} {\bibinfo  {journal} {Phys. Rev. Lett.}\ }\textbf {\bibinfo
  {volume} {120}},\ \bibinfo {pages} {146601} (\bibinfo {year}
  {2018})}\BibitemShut {NoStop}%
\bibitem [{\citenamefont {Jin}\ and\ \citenamefont {Song}(2018)}]{Jin2018}%
  \BibitemOpen
  \bibfield  {author} {\bibinfo {author} {\bibfnamefont {L.}~\bibnamefont
  {Jin}}\ and\ \bibinfo {author} {\bibfnamefont {Z.}~\bibnamefont {Song}},\
  }\href {\doibase 10.1103/PhysRevLett.121.073901} {\bibfield  {journal}
  {\bibinfo  {journal} {Phys. Rev. Lett.}\ }\textbf {\bibinfo {volume} {121}},\
  \bibinfo {pages} {073901} (\bibinfo {year} {2018})}\BibitemShut {NoStop}%
\bibitem [{\citenamefont {Carlstr\"om}\ and\ \citenamefont
  {Bergholtz}(2018)}]{Carlstrom2018}%
  \BibitemOpen
  \bibfield  {author} {\bibinfo {author} {\bibfnamefont {J.}~\bibnamefont
  {Carlstr\"om}}\ and\ \bibinfo {author} {\bibfnamefont {E.~J.}\ \bibnamefont
  {Bergholtz}},\ }\href {\doibase 10.1103/PhysRevA.98.042114} {\bibfield
  {journal} {\bibinfo  {journal} {Phys. Rev. A}\ }\textbf {\bibinfo {volume}
  {98}},\ \bibinfo {pages} {042114} (\bibinfo {year} {2018})}\BibitemShut
  {NoStop}%
\bibitem [{\citenamefont {Pan}\ \emph {et~al.}(2019{\natexlab{a}})\citenamefont
  {Pan}, \citenamefont {Chen},\ and\ \citenamefont {Cui}}]{Pan2019}%
  \BibitemOpen
  \bibfield  {author} {\bibinfo {author} {\bibfnamefont {L.}~\bibnamefont
  {Pan}}, \bibinfo {author} {\bibfnamefont {S.}~\bibnamefont {Chen}}, \ and\
  \bibinfo {author} {\bibfnamefont {X.}~\bibnamefont {Cui}},\ }\href {\doibase
  10.1103/PhysRevA.99.011601} {\bibfield  {journal} {\bibinfo  {journal} {Phys.
  Rev. A}\ }\textbf {\bibinfo {volume} {99}},\ \bibinfo {pages} {011601(R)}
  (\bibinfo {year} {2019}{\natexlab{a}})}\BibitemShut {NoStop}%
\bibitem [{\citenamefont {Moors}\ \emph {et~al.}(2019)\citenamefont {Moors},
  \citenamefont {Zyuzin}, \citenamefont {Zyuzin}, \citenamefont {Tiwari},\ and\
  \citenamefont {Schmidt}}]{Moors2019}%
  \BibitemOpen
  \bibfield  {author} {\bibinfo {author} {\bibfnamefont {K.}~\bibnamefont
  {Moors}}, \bibinfo {author} {\bibfnamefont {A.~A.}\ \bibnamefont {Zyuzin}},
  \bibinfo {author} {\bibfnamefont {A.~Y.}\ \bibnamefont {Zyuzin}}, \bibinfo
  {author} {\bibfnamefont {R.~P.}\ \bibnamefont {Tiwari}}, \ and\ \bibinfo
  {author} {\bibfnamefont {T.~L.}\ \bibnamefont {Schmidt}},\ }\href {\doibase
  10.1103/PhysRevB.99.041116} {\bibfield  {journal} {\bibinfo  {journal} {Phys.
  Rev. B}\ }\textbf {\bibinfo {volume} {99}},\ \bibinfo {pages} {041116(R)}
  (\bibinfo {year} {2019})}\BibitemShut {NoStop}%
\bibitem [{\citenamefont {Budich}\ \emph {et~al.}(2019)\citenamefont {Budich},
  \citenamefont {Carlstr\"om}, \citenamefont {Kunst},\ and\ \citenamefont
  {Bergholtz}}]{Budich2019}%
  \BibitemOpen
  \bibfield  {author} {\bibinfo {author} {\bibfnamefont {J.~C.}\ \bibnamefont
  {Budich}}, \bibinfo {author} {\bibfnamefont {J.}~\bibnamefont {Carlstr\"om}},
  \bibinfo {author} {\bibfnamefont {F.~K.}\ \bibnamefont {Kunst}}, \ and\
  \bibinfo {author} {\bibfnamefont {E.~J.}\ \bibnamefont {Bergholtz}},\ }\href
  {\doibase 10.1103/PhysRevB.99.041406} {\bibfield  {journal} {\bibinfo
  {journal} {Phys. Rev. B}\ }\textbf {\bibinfo {volume} {99}},\ \bibinfo
  {pages} {041406(R)} (\bibinfo {year} {2019})}\BibitemShut {NoStop}%
\bibitem [{\citenamefont {Yang}\ and\ \citenamefont {Hu}(2019)}]{Yang2019}%
  \BibitemOpen
  \bibfield  {author} {\bibinfo {author} {\bibfnamefont {Z.}~\bibnamefont
  {Yang}}\ and\ \bibinfo {author} {\bibfnamefont {J.}~\bibnamefont {Hu}},\
  }\href {\doibase 10.1103/PhysRevB.99.081102} {\bibfield  {journal} {\bibinfo
  {journal} {Phys. Rev. B}\ }\textbf {\bibinfo {volume} {99}},\ \bibinfo
  {pages} {081102(R)} (\bibinfo {year} {2019})}\BibitemShut {NoStop}%
\bibitem [{\citenamefont {Zhou}\ \emph {et~al.}(2019)\citenamefont {Zhou},
  \citenamefont {Lee}, \citenamefont {Liu},\ and\ \citenamefont
  {Zhen}}]{Zhou2019}%
  \BibitemOpen
  \bibfield  {author} {\bibinfo {author} {\bibfnamefont {H.}~\bibnamefont
  {Zhou}}, \bibinfo {author} {\bibfnamefont {J.~Y.}\ \bibnamefont {Lee}},
  \bibinfo {author} {\bibfnamefont {S.}~\bibnamefont {Liu}}, \ and\ \bibinfo
  {author} {\bibfnamefont {B.}~\bibnamefont {Zhen}},\ }\href {\doibase
  10.1364/OPTICA.6.000190} {\bibfield  {journal} {\bibinfo  {journal} {Optica}\
  }\textbf {\bibinfo {volume} {6}},\ \bibinfo {pages} {190} (\bibinfo {year}
  {2019})}\BibitemShut {NoStop}%
\bibitem [{\citenamefont {Zhu}\ \emph {et~al.}(2019)\citenamefont {Zhu},
  \citenamefont {Wang},\ and\ \citenamefont {Chong}}]{Zhu2019}%
  \BibitemOpen
  \bibfield  {author} {\bibinfo {author} {\bibfnamefont {B.}~\bibnamefont
  {Zhu}}, \bibinfo {author} {\bibfnamefont {Q.~J.}\ \bibnamefont {Wang}}, \
  and\ \bibinfo {author} {\bibfnamefont {Y.~D.}\ \bibnamefont {Chong}},\ }\href
  {\doibase 10.1103/PhysRevA.99.033829} {\bibfield  {journal} {\bibinfo
  {journal} {Phys. Rev. A}\ }\textbf {\bibinfo {volume} {99}},\ \bibinfo
  {pages} {033829} (\bibinfo {year} {2019})}\BibitemShut {NoStop}%
\bibitem [{\citenamefont {Hatano}(2019)}]{Hatano2019}%
  \BibitemOpen
  \bibfield  {author} {\bibinfo {author} {\bibfnamefont {N.}~\bibnamefont
  {Hatano}},\ }\href@noop {} {\bibfield  {journal} {\bibinfo  {journal} {Mol.
  Phys.}\ }\textbf {\bibinfo {volume} {117}},\ \bibinfo {pages} {2121}
  (\bibinfo {year} {2019})}\BibitemShut {NoStop}%
\bibitem [{\citenamefont {Chen}\ \emph
  {et~al.}(2019{\natexlab{a}})\citenamefont {Chen}, \citenamefont {Zhang},
  \citenamefont {Yang}, \citenamefont {Wu},\ and\ \citenamefont
  {Zhang}}]{Chen2019}%
  \BibitemOpen
  \bibfield  {author} {\bibinfo {author} {\bibfnamefont {S.}~\bibnamefont
  {Chen}}, \bibinfo {author} {\bibfnamefont {W.}~\bibnamefont {Zhang}},
  \bibinfo {author} {\bibfnamefont {B.}~\bibnamefont {Yang}}, \bibinfo {author}
  {\bibfnamefont {T.}~\bibnamefont {Wu}}, \ and\ \bibinfo {author}
  {\bibfnamefont {X.}~\bibnamefont {Zhang}},\ }\href@noop {} {\bibfield
  {journal} {\bibinfo  {journal} {Sci. Rep.}\ }\textbf {\bibinfo {volume}
  {9}},\ \bibinfo {pages} {5551} (\bibinfo {year}
  {2019}{\natexlab{a}})}\BibitemShut {NoStop}%
\bibitem [{\citenamefont {Zhang}\ \emph
  {et~al.}(2019{\natexlab{b}})\citenamefont {Zhang}, \citenamefont {Wang},\
  and\ \citenamefont {Song}}]{Zhang2019}%
  \BibitemOpen
  \bibfield  {author} {\bibinfo {author} {\bibfnamefont {K.~L.}\ \bibnamefont
  {Zhang}}, \bibinfo {author} {\bibfnamefont {P.}~\bibnamefont {Wang}}, \ and\
  \bibinfo {author} {\bibfnamefont {Z.}~\bibnamefont {Song}},\ }\href {\doibase
  10.1103/PhysRevA.99.042111} {\bibfield  {journal} {\bibinfo  {journal} {Phys.
  Rev. A}\ }\textbf {\bibinfo {volume} {99}},\ \bibinfo {pages} {042111}
  (\bibinfo {year} {2019}{\natexlab{b}})}\BibitemShut {NoStop}%
\bibitem [{\citenamefont {Zhong}\ \emph {et~al.}(2019)\citenamefont {Zhong},
  \citenamefont {Ren}, \citenamefont {Khajavikhan}, \citenamefont
  {Christodoulides}, \citenamefont {\"Ozdemir},\ and\ \citenamefont
  {El-Ganainy}}]{Zhong2019}%
  \BibitemOpen
  \bibfield  {author} {\bibinfo {author} {\bibfnamefont {Q.}~\bibnamefont
  {Zhong}}, \bibinfo {author} {\bibfnamefont {J.}~\bibnamefont {Ren}}, \bibinfo
  {author} {\bibfnamefont {M.}~\bibnamefont {Khajavikhan}}, \bibinfo {author}
  {\bibfnamefont {D.~N.}\ \bibnamefont {Christodoulides}}, \bibinfo {author}
  {\bibfnamefont {{\c S}.~K.}\ \bibnamefont {\"Ozdemir}}, \ and\ \bibinfo
  {author} {\bibfnamefont {R.}~\bibnamefont {El-Ganainy}},\ }\href {\doibase
  10.1103/PhysRevLett.122.153902} {\bibfield  {journal} {\bibinfo  {journal}
  {Phys. Rev. Lett.}\ }\textbf {\bibinfo {volume} {122}},\ \bibinfo {pages}
  {153902} (\bibinfo {year} {2019})}\BibitemShut {NoStop}%
\bibitem [{\citenamefont {Carlstr\"om}\ \emph {et~al.}(2019)\citenamefont
  {Carlstr\"om}, \citenamefont {St\aa{}lhammar}, \citenamefont {Budich},\ and\
  \citenamefont {Bergholtz}}]{Carlstrom2019}%
  \BibitemOpen
  \bibfield  {author} {\bibinfo {author} {\bibfnamefont {J.}~\bibnamefont
  {Carlstr\"om}}, \bibinfo {author} {\bibfnamefont {M.}~\bibnamefont
  {St\aa{}lhammar}}, \bibinfo {author} {\bibfnamefont {J.~C.}\ \bibnamefont
  {Budich}}, \ and\ \bibinfo {author} {\bibfnamefont {E.~J.}\ \bibnamefont
  {Bergholtz}},\ }\href {\doibase 10.1103/PhysRevB.99.161115} {\bibfield
  {journal} {\bibinfo  {journal} {Phys. Rev. B}\ }\textbf {\bibinfo {volume}
  {99}},\ \bibinfo {pages} {161115} (\bibinfo {year} {2019})}\BibitemShut
  {NoStop}%
\bibitem [{\citenamefont {Zyuzin}\ and\ \citenamefont
  {Simon}(2019)}]{Zyuzin2019}%
  \BibitemOpen
  \bibfield  {author} {\bibinfo {author} {\bibfnamefont {A.~A.}\ \bibnamefont
  {Zyuzin}}\ and\ \bibinfo {author} {\bibfnamefont {P.}~\bibnamefont {Simon}},\
  }\href {\doibase 10.1103/PhysRevB.99.165145} {\bibfield  {journal} {\bibinfo
  {journal} {Phys. Rev. B}\ }\textbf {\bibinfo {volume} {99}},\ \bibinfo
  {pages} {165145} (\bibinfo {year} {2019})}\BibitemShut {NoStop}%
\bibitem [{\citenamefont {Xiao}\ \emph
  {et~al.}(2019{\natexlab{b}})\citenamefont {Xiao}, \citenamefont {Zhang},
  \citenamefont {Hang},\ and\ \citenamefont {Chan}}]{Xiao2019}%
  \BibitemOpen
  \bibfield  {author} {\bibinfo {author} {\bibfnamefont {Y.-X.}\ \bibnamefont
  {Xiao}}, \bibinfo {author} {\bibfnamefont {Z.-Q.}\ \bibnamefont {Zhang}},
  \bibinfo {author} {\bibfnamefont {Z.~H.}\ \bibnamefont {Hang}}, \ and\
  \bibinfo {author} {\bibfnamefont {C.~T.}\ \bibnamefont {Chan}},\ }\href
  {\doibase 10.1103/PhysRevB.99.241403} {\bibfield  {journal} {\bibinfo
  {journal} {Phys. Rev. B}\ }\textbf {\bibinfo {volume} {99}},\ \bibinfo
  {pages} {241403(R)} (\bibinfo {year} {2019}{\natexlab{b}})}\BibitemShut
  {NoStop}%
\bibitem [{\citenamefont {Pan}\ \emph {et~al.}(2019{\natexlab{b}})\citenamefont
  {Pan}, \citenamefont {Chen},\ and\ \citenamefont {Cui}}]{Pan2019v2}%
  \BibitemOpen
  \bibfield  {author} {\bibinfo {author} {\bibfnamefont {L.}~\bibnamefont
  {Pan}}, \bibinfo {author} {\bibfnamefont {S.}~\bibnamefont {Chen}}, \ and\
  \bibinfo {author} {\bibfnamefont {X.}~\bibnamefont {Cui}},\ }\href {\doibase
  10.1103/PhysRevA.99.063616} {\bibfield  {journal} {\bibinfo  {journal} {Phys.
  Rev. A}\ }\textbf {\bibinfo {volume} {99}},\ \bibinfo {pages} {063616}
  (\bibinfo {year} {2019}{\natexlab{b}})}\BibitemShut {NoStop}%
\bibitem [{\citenamefont {Rui}\ \emph {et~al.}(2019)\citenamefont {Rui},
  \citenamefont {Zhao},\ and\ \citenamefont {Schnyder}}]{Rui2019}%
  \BibitemOpen
  \bibfield  {author} {\bibinfo {author} {\bibfnamefont {W.~B.}\ \bibnamefont
  {Rui}}, \bibinfo {author} {\bibfnamefont {Y.~X.}\ \bibnamefont {Zhao}}, \
  and\ \bibinfo {author} {\bibfnamefont {A.~P.}\ \bibnamefont {Schnyder}},\
  }\href {\doibase 10.1103/PhysRevB.99.241110} {\bibfield  {journal} {\bibinfo
  {journal} {Phys. Rev. B}\ }\textbf {\bibinfo {volume} {99}},\ \bibinfo
  {pages} {241110} (\bibinfo {year} {2019})}\BibitemShut {NoStop}%
\bibitem [{\citenamefont {Kimura}\ \emph {et~al.}(2019)\citenamefont {Kimura},
  \citenamefont {Yoshida},\ and\ \citenamefont {Kawakami}}]{Kimura2019}%
  \BibitemOpen
  \bibfield  {author} {\bibinfo {author} {\bibfnamefont {K.}~\bibnamefont
  {Kimura}}, \bibinfo {author} {\bibfnamefont {T.}~\bibnamefont {Yoshida}}, \
  and\ \bibinfo {author} {\bibfnamefont {N.}~\bibnamefont {Kawakami}},\ }\href
  {\doibase 10.1103/PhysRevB.100.115124} {\bibfield  {journal} {\bibinfo
  {journal} {Phys. Rev. B}\ }\textbf {\bibinfo {volume} {100}},\ \bibinfo
  {pages} {115124} (\bibinfo {year} {2019})}\BibitemShut {NoStop}%
\bibitem [{\citenamefont {Okugawa}\ and\ \citenamefont
  {Yokoyama}(2019)}]{Okugawa2019}%
  \BibitemOpen
  \bibfield  {author} {\bibinfo {author} {\bibfnamefont {R.}~\bibnamefont
  {Okugawa}}\ and\ \bibinfo {author} {\bibfnamefont {T.}~\bibnamefont
  {Yokoyama}},\ }\href {\doibase 10.1103/PhysRevB.99.041202} {\bibfield
  {journal} {\bibinfo  {journal} {Phys. Rev. B}\ }\textbf {\bibinfo {volume}
  {99}},\ \bibinfo {pages} {041202(R)} (\bibinfo {year} {2019})}\BibitemShut
  {NoStop}%
\bibitem [{\citenamefont {Kawabata}\ \emph
  {et~al.}(2019{\natexlab{a}})\citenamefont {Kawabata}, \citenamefont
  {Bessho},\ and\ \citenamefont {Sato}}]{Kawabata2019v2}%
  \BibitemOpen
  \bibfield  {author} {\bibinfo {author} {\bibfnamefont {K.}~\bibnamefont
  {Kawabata}}, \bibinfo {author} {\bibfnamefont {T.}~\bibnamefont {Bessho}}, \
  and\ \bibinfo {author} {\bibfnamefont {M.}~\bibnamefont {Sato}},\ }\href
  {\doibase 10.1103/PhysRevLett.123.066405} {\bibfield  {journal} {\bibinfo
  {journal} {Phys. Rev. Lett.}\ }\textbf {\bibinfo {volume} {123}},\ \bibinfo
  {pages} {066405} (\bibinfo {year} {2019}{\natexlab{a}})}\BibitemShut
  {NoStop}%
\bibitem [{\citenamefont {Yoshida}\ \emph {et~al.}(2019)\citenamefont
  {Yoshida}, \citenamefont {Peters}, \citenamefont {Kawakami},\ and\
  \citenamefont {Hatsugai}}]{Yoshida2019}%
  \BibitemOpen
  \bibfield  {author} {\bibinfo {author} {\bibfnamefont {T.}~\bibnamefont
  {Yoshida}}, \bibinfo {author} {\bibfnamefont {R.}~\bibnamefont {Peters}},
  \bibinfo {author} {\bibfnamefont {N.}~\bibnamefont {Kawakami}}, \ and\
  \bibinfo {author} {\bibfnamefont {Y.}~\bibnamefont {Hatsugai}},\ }\href
  {\doibase 10.1103/PhysRevB.99.121101} {\bibfield  {journal} {\bibinfo
  {journal} {Phys. Rev. B}\ }\textbf {\bibinfo {volume} {99}},\ \bibinfo
  {pages} {121101(R)} (\bibinfo {year} {2019})}\BibitemShut {NoStop}%
\bibitem [{\citenamefont {Lin}\ \emph {et~al.}(2019)\citenamefont {Lin},
  \citenamefont {Jin},\ and\ \citenamefont {Song}}]{Lin2019}%
  \BibitemOpen
  \bibfield  {author} {\bibinfo {author} {\bibfnamefont {S.}~\bibnamefont
  {Lin}}, \bibinfo {author} {\bibfnamefont {L.}~\bibnamefont {Jin}}, \ and\
  \bibinfo {author} {\bibfnamefont {Z.}~\bibnamefont {Song}},\ }\href {\doibase
  10.1103/PhysRevB.99.165148} {\bibfield  {journal} {\bibinfo  {journal} {Phys.
  Rev. B}\ }\textbf {\bibinfo {volume} {99}},\ \bibinfo {pages} {165148}
  (\bibinfo {year} {2019})}\BibitemShut {NoStop}%
\bibitem [{\citenamefont {Yoshida}\ and\ \citenamefont
  {Hatsugai}(2019)}]{Yoshida2019v2}%
  \BibitemOpen
  \bibfield  {author} {\bibinfo {author} {\bibfnamefont {T.}~\bibnamefont
  {Yoshida}}\ and\ \bibinfo {author} {\bibfnamefont {Y.}~\bibnamefont
  {Hatsugai}},\ }\href {\doibase 10.1103/PhysRevB.100.054109} {\bibfield
  {journal} {\bibinfo  {journal} {Phys. Rev. B}\ }\textbf {\bibinfo {volume}
  {100}},\ \bibinfo {pages} {054109} (\bibinfo {year} {2019})}\BibitemShut
  {NoStop}%
\bibitem [{\citenamefont {Hu}\ and\ \citenamefont {Hughes}(2011)}]{Hu2011}%
  \BibitemOpen
  \bibfield  {author} {\bibinfo {author} {\bibfnamefont {Y.~C.}\ \bibnamefont
  {Hu}}\ and\ \bibinfo {author} {\bibfnamefont {T.~L.}\ \bibnamefont
  {Hughes}},\ }\href {\doibase 10.1103/PhysRevB.84.153101} {\bibfield
  {journal} {\bibinfo  {journal} {Phys. Rev. B}\ }\textbf {\bibinfo {volume}
  {84}},\ \bibinfo {pages} {153101} (\bibinfo {year} {2011})}\BibitemShut
  {NoStop}%
\bibitem [{\citenamefont {Esaki}\ \emph {et~al.}(2011)\citenamefont {Esaki},
  \citenamefont {Sato}, \citenamefont {Hasebe},\ and\ \citenamefont
  {Kohmoto}}]{Esaki2011}%
  \BibitemOpen
  \bibfield  {author} {\bibinfo {author} {\bibfnamefont {K.}~\bibnamefont
  {Esaki}}, \bibinfo {author} {\bibfnamefont {M.}~\bibnamefont {Sato}},
  \bibinfo {author} {\bibfnamefont {K.}~\bibnamefont {Hasebe}}, \ and\ \bibinfo
  {author} {\bibfnamefont {M.}~\bibnamefont {Kohmoto}},\ }\href {\doibase
  10.1103/PhysRevB.84.205128} {\bibfield  {journal} {\bibinfo  {journal} {Phys.
  Rev. B}\ }\textbf {\bibinfo {volume} {84}},\ \bibinfo {pages} {205128}
  (\bibinfo {year} {2011})}\BibitemShut {NoStop}%
\bibitem [{\citenamefont {Kozii}\ and\ \citenamefont {Fu}()}]{Kozii2017}%
  \BibitemOpen
  \bibfield  {author} {\bibinfo {author} {\bibfnamefont {V.}~\bibnamefont
  {Kozii}}\ and\ \bibinfo {author} {\bibfnamefont {L.}~\bibnamefont {Fu}},\
  }\href@noop {} {\bibinfo  {journal} {arXiv:1708.05841}\ }\BibitemShut
  {NoStop}%
\bibitem [{\citenamefont {Harari}\ \emph {et~al.}(2018)\citenamefont {Harari},
  \citenamefont {Bandres}, \citenamefont {Lumer}, \citenamefont {Rechtsman},
  \citenamefont {Chong}, \citenamefont {Khajavikhan}, \citenamefont
  {Christodoulides},\ and\ \citenamefont {Segev}}]{Harari2018}%
  \BibitemOpen
\bibfield  {journal} {  }\bibfield  {author} {\bibinfo {author} {\bibfnamefont
  {G.}~\bibnamefont {Harari}}, \bibinfo {author} {\bibfnamefont {M.~A.}\
  \bibnamefont {Bandres}}, \bibinfo {author} {\bibfnamefont {Y.}~\bibnamefont
  {Lumer}}, \bibinfo {author} {\bibfnamefont {M.~C.}\ \bibnamefont
  {Rechtsman}}, \bibinfo {author} {\bibfnamefont {Y.~D.}\ \bibnamefont
  {Chong}}, \bibinfo {author} {\bibfnamefont {M.}~\bibnamefont {Khajavikhan}},
  \bibinfo {author} {\bibfnamefont {D.~N.}\ \bibnamefont {Christodoulides}}, \
  and\ \bibinfo {author} {\bibfnamefont {M.}~\bibnamefont {Segev}},\ }\href
  {\doibase 10.1126/science.aar4003} {\bibfield  {journal} {\bibinfo  {journal}
  {Science}\ }\textbf {\bibinfo {volume} {359}},\ \bibinfo {pages} {eaar4003}
  (\bibinfo {year} {2018})}\BibitemShut {NoStop}%
\bibitem [{\citenamefont {Shen}\ \emph {et~al.}(2018)\citenamefont {Shen},
  \citenamefont {Zhen},\ and\ \citenamefont {Fu}}]{Shen2018}%
  \BibitemOpen
  \bibfield  {author} {\bibinfo {author} {\bibfnamefont {H.}~\bibnamefont
  {Shen}}, \bibinfo {author} {\bibfnamefont {B.}~\bibnamefont {Zhen}}, \ and\
  \bibinfo {author} {\bibfnamefont {L.}~\bibnamefont {Fu}},\ }\href {\doibase
  10.1103/PhysRevLett.120.146402} {\bibfield  {journal} {\bibinfo  {journal}
  {Phys. Rev. Lett.}\ }\textbf {\bibinfo {volume} {120}},\ \bibinfo {pages}
  {146402} (\bibinfo {year} {2018})}\BibitemShut {NoStop}%
\bibitem [{\citenamefont {Yoshida}\ \emph {et~al.}(2018)\citenamefont
  {Yoshida}, \citenamefont {Peters},\ and\ \citenamefont
  {Kawakami}}]{Yoshida2018}%
  \BibitemOpen
  \bibfield  {author} {\bibinfo {author} {\bibfnamefont {T.}~\bibnamefont
  {Yoshida}}, \bibinfo {author} {\bibfnamefont {R.}~\bibnamefont {Peters}}, \
  and\ \bibinfo {author} {\bibfnamefont {N.}~\bibnamefont {Kawakami}},\ }\href
  {\doibase 10.1103/PhysRevB.98.035141} {\bibfield  {journal} {\bibinfo
  {journal} {Phys. Rev. B}\ }\textbf {\bibinfo {volume} {98}},\ \bibinfo
  {pages} {035141} (\bibinfo {year} {2018})}\BibitemShut {NoStop}%
\bibitem [{\citenamefont {Gong}\ \emph {et~al.}(2018)\citenamefont {Gong},
  \citenamefont {Ashida}, \citenamefont {Kawabata}, \citenamefont {Takasan},
  \citenamefont {Higashikawa},\ and\ \citenamefont {Ueda}}]{Gong2018}%
  \BibitemOpen
  \bibfield  {author} {\bibinfo {author} {\bibfnamefont {Z.}~\bibnamefont
  {Gong}}, \bibinfo {author} {\bibfnamefont {Y.}~\bibnamefont {Ashida}},
  \bibinfo {author} {\bibfnamefont {K.}~\bibnamefont {Kawabata}}, \bibinfo
  {author} {\bibfnamefont {K.}~\bibnamefont {Takasan}}, \bibinfo {author}
  {\bibfnamefont {S.}~\bibnamefont {Higashikawa}}, \ and\ \bibinfo {author}
  {\bibfnamefont {M.}~\bibnamefont {Ueda}},\ }\href {\doibase
  10.1103/PhysRevX.8.031079} {\bibfield  {journal} {\bibinfo  {journal} {Phys.
  Rev. X}\ }\textbf {\bibinfo {volume} {8}},\ \bibinfo {pages} {031079}
  (\bibinfo {year} {2018})}\BibitemShut {NoStop}%
\bibitem [{\citenamefont {Philip}\ \emph {et~al.}(2018)\citenamefont {Philip},
  \citenamefont {Hirsbrunner},\ and\ \citenamefont {Gilbert}}]{Philip2018}%
  \BibitemOpen
  \bibfield  {author} {\bibinfo {author} {\bibfnamefont {T.~M.}\ \bibnamefont
  {Philip}}, \bibinfo {author} {\bibfnamefont {M.~R.}\ \bibnamefont
  {Hirsbrunner}}, \ and\ \bibinfo {author} {\bibfnamefont {M.~J.}\ \bibnamefont
  {Gilbert}},\ }\href {\doibase 10.1103/PhysRevB.98.155430} {\bibfield
  {journal} {\bibinfo  {journal} {Phys. Rev. B}\ }\textbf {\bibinfo {volume}
  {98}},\ \bibinfo {pages} {155430} (\bibinfo {year} {2018})}\BibitemShut
  {NoStop}%
\bibitem [{\citenamefont {Chen}\ and\ \citenamefont {Zhai}(2018)}]{Chen2018}%
  \BibitemOpen
  \bibfield  {author} {\bibinfo {author} {\bibfnamefont {Y.}~\bibnamefont
  {Chen}}\ and\ \bibinfo {author} {\bibfnamefont {H.}~\bibnamefont {Zhai}},\
  }\href {\doibase 10.1103/PhysRevB.98.245130} {\bibfield  {journal} {\bibinfo
  {journal} {Phys. Rev. B}\ }\textbf {\bibinfo {volume} {98}},\ \bibinfo
  {pages} {245130} (\bibinfo {year} {2018})}\BibitemShut {NoStop}%
\bibitem [{\citenamefont {Liu}\ \emph {et~al.}(2019{\natexlab{a}})\citenamefont
  {Liu}, \citenamefont {Jiang},\ and\ \citenamefont {Chen}}]{Liu2019}%
  \BibitemOpen
  \bibfield  {author} {\bibinfo {author} {\bibfnamefont {C.-H.}\ \bibnamefont
  {Liu}}, \bibinfo {author} {\bibfnamefont {H.}~\bibnamefont {Jiang}}, \ and\
  \bibinfo {author} {\bibfnamefont {S.}~\bibnamefont {Chen}},\ }\href {\doibase
  10.1103/PhysRevB.99.125103} {\bibfield  {journal} {\bibinfo  {journal} {Phys.
  Rev. B}\ }\textbf {\bibinfo {volume} {99}},\ \bibinfo {pages} {125103}
  (\bibinfo {year} {2019}{\natexlab{a}})}\BibitemShut {NoStop}%
\bibitem [{\citenamefont {Papaj}\ \emph {et~al.}(2019)\citenamefont {Papaj},
  \citenamefont {Isobe},\ and\ \citenamefont {Fu}}]{Papaj2019}%
  \BibitemOpen
  \bibfield  {author} {\bibinfo {author} {\bibfnamefont {M.}~\bibnamefont
  {Papaj}}, \bibinfo {author} {\bibfnamefont {H.}~\bibnamefont {Isobe}}, \ and\
  \bibinfo {author} {\bibfnamefont {L.}~\bibnamefont {Fu}},\ }\href {\doibase
  10.1103/PhysRevB.99.201107} {\bibfield  {journal} {\bibinfo  {journal} {Phys.
  Rev. B}\ }\textbf {\bibinfo {volume} {99}},\ \bibinfo {pages} {201107}
  (\bibinfo {year} {2019})}\BibitemShut {NoStop}%
\bibitem [{\citenamefont {Ghatak}\ and\ \citenamefont
  {Das}(2019)}]{Ghatak2019t}%
  \BibitemOpen
  \bibfield  {author} {\bibinfo {author} {\bibfnamefont {A.}~\bibnamefont
  {Ghatak}}\ and\ \bibinfo {author} {\bibfnamefont {T.}~\bibnamefont {Das}},\
  }\href@noop {} {\bibfield  {journal} {\bibinfo  {journal} {J. Phys.: Condens.
  Matter}\ }\textbf {\bibinfo {volume} {31}},\ \bibinfo {pages} {263001}
  (\bibinfo {year} {2019})}\BibitemShut {NoStop}%
\bibitem [{\citenamefont {McClarty}\ and\ \citenamefont
  {Rau}(2019)}]{McClarty2019}%
  \BibitemOpen
  \bibfield  {author} {\bibinfo {author} {\bibfnamefont {P.~A.}\ \bibnamefont
  {McClarty}}\ and\ \bibinfo {author} {\bibfnamefont {J.~G.}\ \bibnamefont
  {Rau}},\ }\href {\doibase 10.1103/PhysRevB.100.100405} {\bibfield  {journal}
  {\bibinfo  {journal} {Phys. Rev. B}\ }\textbf {\bibinfo {volume} {100}},\
  \bibinfo {pages} {100405(R)} (\bibinfo {year} {2019})}\BibitemShut {NoStop}%
\bibitem [{\citenamefont {Xiong}(2018)}]{Ye2018}%
  \BibitemOpen
  \bibfield  {author} {\bibinfo {author} {\bibfnamefont {Y.}~\bibnamefont
  {Xiong}},\ }\href@noop {} {\bibfield  {journal} {\bibinfo  {journal} {J.
  Phys. Commun.}\ }\textbf {\bibinfo {volume} {2}},\ \bibinfo {pages} {035043}
  (\bibinfo {year} {2018})}\BibitemShut {NoStop}%
\bibitem [{\citenamefont {Malzard}\ and\ \citenamefont
  {Schomerus}(2018)}]{Malzard2018}%
  \BibitemOpen
  \bibfield  {author} {\bibinfo {author} {\bibfnamefont {S.}~\bibnamefont
  {Malzard}}\ and\ \bibinfo {author} {\bibfnamefont {H.}~\bibnamefont
  {Schomerus}},\ }\href {\doibase 10.1103/PhysRevA.98.033807} {\bibfield
  {journal} {\bibinfo  {journal} {Phys. Rev. A}\ }\textbf {\bibinfo {volume}
  {98}},\ \bibinfo {pages} {033807} (\bibinfo {year} {2018})}\BibitemShut
  {NoStop}%
\bibitem [{\citenamefont {Yao}\ \emph {et~al.}(2018)\citenamefont {Yao},
  \citenamefont {Song},\ and\ \citenamefont {Wang}}]{Yao2018v2}%
  \BibitemOpen
  \bibfield  {author} {\bibinfo {author} {\bibfnamefont {S.}~\bibnamefont
  {Yao}}, \bibinfo {author} {\bibfnamefont {F.}~\bibnamefont {Song}}, \ and\
  \bibinfo {author} {\bibfnamefont {Z.}~\bibnamefont {Wang}},\ }\href {\doibase
  10.1103/PhysRevLett.121.136802} {\bibfield  {journal} {\bibinfo  {journal}
  {Phys. Rev. Lett.}\ }\textbf {\bibinfo {volume} {121}},\ \bibinfo {pages}
  {136802} (\bibinfo {year} {2018})}\BibitemShut {NoStop}%
\bibitem [{\citenamefont {Kawabata}\ \emph {et~al.}(2018)\citenamefont
  {Kawabata}, \citenamefont {Shiozaki},\ and\ \citenamefont
  {Ueda}}]{Kawabata2018v2}%
  \BibitemOpen
  \bibfield  {author} {\bibinfo {author} {\bibfnamefont {K.}~\bibnamefont
  {Kawabata}}, \bibinfo {author} {\bibfnamefont {K.}~\bibnamefont {Shiozaki}},
  \ and\ \bibinfo {author} {\bibfnamefont {M.}~\bibnamefont {Ueda}},\ }\href
  {\doibase 10.1103/PhysRevB.98.165148} {\bibfield  {journal} {\bibinfo
  {journal} {Phys. Rev. B}\ }\textbf {\bibinfo {volume} {98}},\ \bibinfo
  {pages} {165148} (\bibinfo {year} {2018})}\BibitemShut {NoStop}%
\bibitem [{\citenamefont {Takata}\ and\ \citenamefont
  {Notomi}(2018)}]{Takata2018}%
  \BibitemOpen
  \bibfield  {author} {\bibinfo {author} {\bibfnamefont {K.}~\bibnamefont
  {Takata}}\ and\ \bibinfo {author} {\bibfnamefont {M.}~\bibnamefont
  {Notomi}},\ }\href {\doibase 10.1103/PhysRevLett.121.213902} {\bibfield
  {journal} {\bibinfo  {journal} {Phys. Rev. Lett.}\ }\textbf {\bibinfo
  {volume} {121}},\ \bibinfo {pages} {213902} (\bibinfo {year}
  {2018})}\BibitemShut {NoStop}%
\bibitem [{\citenamefont {Kawabata}\ \emph
  {et~al.}(2019{\natexlab{b}})\citenamefont {Kawabata}, \citenamefont
  {Higashikawa}, \citenamefont {Gong}, \citenamefont {Ashida},\ and\
  \citenamefont {Ueda}}]{Kawabata2019}%
  \BibitemOpen
  \bibfield  {author} {\bibinfo {author} {\bibfnamefont {K.}~\bibnamefont
  {Kawabata}}, \bibinfo {author} {\bibfnamefont {S.}~\bibnamefont
  {Higashikawa}}, \bibinfo {author} {\bibfnamefont {Z.}~\bibnamefont {Gong}},
  \bibinfo {author} {\bibfnamefont {Y.}~\bibnamefont {Ashida}}, \ and\ \bibinfo
  {author} {\bibfnamefont {M.}~\bibnamefont {Ueda}},\ }\href@noop {} {\bibfield
   {journal} {\bibinfo  {journal} {Nat. Commun.}\ }\textbf {\bibinfo {volume}
  {10}},\ \bibinfo {pages} {297} (\bibinfo {year}
  {2019}{\natexlab{b}})}\BibitemShut {NoStop}%
\bibitem [{\citenamefont {Bliokh}\ \emph {et~al.}(2019)\citenamefont {Bliokh},
  \citenamefont {Leykam}, \citenamefont {Lein},\ and\ \citenamefont
  {Nori}}]{Bliokh2019}%
  \BibitemOpen
  \bibfield  {author} {\bibinfo {author} {\bibfnamefont {K.~Y.}\ \bibnamefont
  {Bliokh}}, \bibinfo {author} {\bibfnamefont {D.}~\bibnamefont {Leykam}},
  \bibinfo {author} {\bibfnamefont {M.}~\bibnamefont {Lein}}, \ and\ \bibinfo
  {author} {\bibfnamefont {F.}~\bibnamefont {Nori}},\ }\href@noop {} {\bibfield
   {journal} {\bibinfo  {journal} {Nat. Commun.}\ }\textbf {\bibinfo {volume}
  {10}},\ \bibinfo {pages} {580} (\bibinfo {year} {2019})}\BibitemShut
  {NoStop}%
\bibitem [{\citenamefont {Wang}\ \emph
  {et~al.}(2019{\natexlab{b}})\citenamefont {Wang}, \citenamefont {Ruan},\ and\
  \citenamefont {Zhang}}]{Wang2019}%
  \BibitemOpen
  \bibfield  {author} {\bibinfo {author} {\bibfnamefont {H.}~\bibnamefont
  {Wang}}, \bibinfo {author} {\bibfnamefont {J.}~\bibnamefont {Ruan}}, \ and\
  \bibinfo {author} {\bibfnamefont {H.}~\bibnamefont {Zhang}},\ }\href
  {\doibase 10.1103/PhysRevB.99.075130} {\bibfield  {journal} {\bibinfo
  {journal} {Phys. Rev. B}\ }\textbf {\bibinfo {volume} {99}},\ \bibinfo
  {pages} {075130} (\bibinfo {year} {2019}{\natexlab{b}})}\BibitemShut
  {NoStop}%
\bibitem [{\citenamefont {Liu}\ \emph {et~al.}(2019{\natexlab{b}})\citenamefont
  {Liu}, \citenamefont {Zhang}, \citenamefont {Ai}, \citenamefont {Gong},
  \citenamefont {Kawabata}, \citenamefont {Ueda},\ and\ \citenamefont
  {Nori}}]{tLiu2019}%
  \BibitemOpen
  \bibfield  {author} {\bibinfo {author} {\bibfnamefont {T.}~\bibnamefont
  {Liu}}, \bibinfo {author} {\bibfnamefont {Y.-R.}\ \bibnamefont {Zhang}},
  \bibinfo {author} {\bibfnamefont {Q.}~\bibnamefont {Ai}}, \bibinfo {author}
  {\bibfnamefont {Z.}~\bibnamefont {Gong}}, \bibinfo {author} {\bibfnamefont
  {K.}~\bibnamefont {Kawabata}}, \bibinfo {author} {\bibfnamefont
  {M.}~\bibnamefont {Ueda}}, \ and\ \bibinfo {author} {\bibfnamefont
  {F.}~\bibnamefont {Nori}},\ }\href {\doibase 10.1103/PhysRevLett.122.076801}
  {\bibfield  {journal} {\bibinfo  {journal} {Phys. Rev. Lett.}\ }\textbf
  {\bibinfo {volume} {122}},\ \bibinfo {pages} {076801} (\bibinfo {year}
  {2019}{\natexlab{b}})}\BibitemShut {NoStop}%
\bibitem [{\citenamefont {Edvardsson}\ \emph {et~al.}(2019)\citenamefont
  {Edvardsson}, \citenamefont {Kunst},\ and\ \citenamefont
  {Bergholtz}}]{Edvardsson2019}%
  \BibitemOpen
  \bibfield  {author} {\bibinfo {author} {\bibfnamefont {E.}~\bibnamefont
  {Edvardsson}}, \bibinfo {author} {\bibfnamefont {F.~K.}\ \bibnamefont
  {Kunst}}, \ and\ \bibinfo {author} {\bibfnamefont {E.~J.}\ \bibnamefont
  {Bergholtz}},\ }\href {\doibase 10.1103/PhysRevB.99.081302} {\bibfield
  {journal} {\bibinfo  {journal} {Phys. Rev. B}\ }\textbf {\bibinfo {volume}
  {99}},\ \bibinfo {pages} {081302(R)} (\bibinfo {year} {2019})}\BibitemShut
  {NoStop}%
\bibitem [{\citenamefont {Zhang}\ \emph
  {et~al.}(2019{\natexlab{c}})\citenamefont {Zhang}, \citenamefont {Wu},
  \citenamefont {Jin},\ and\ \citenamefont {Song}}]{KLZhang2019}%
  \BibitemOpen
  \bibfield  {author} {\bibinfo {author} {\bibfnamefont {K.~L.}\ \bibnamefont
  {Zhang}}, \bibinfo {author} {\bibfnamefont {H.~C.}\ \bibnamefont {Wu}},
  \bibinfo {author} {\bibfnamefont {L.}~\bibnamefont {Jin}}, \ and\ \bibinfo
  {author} {\bibfnamefont {Z.}~\bibnamefont {Song}},\ }\href {\doibase
  10.1103/PhysRevB.100.045141} {\bibfield  {journal} {\bibinfo  {journal}
  {Phys. Rev. B}\ }\textbf {\bibinfo {volume} {100}},\ \bibinfo {pages}
  {045141} (\bibinfo {year} {2019}{\natexlab{c}})}\BibitemShut {NoStop}%
\bibitem [{\citenamefont {Lieu}(2019)}]{Lieu2019}%
  \BibitemOpen
  \bibfield  {author} {\bibinfo {author} {\bibfnamefont {S.}~\bibnamefont
  {Lieu}},\ }\href {\doibase 10.1103/PhysRevB.100.085110} {\bibfield  {journal}
  {\bibinfo  {journal} {Phys. Rev. B}\ }\textbf {\bibinfo {volume} {100}},\
  \bibinfo {pages} {085110} (\bibinfo {year} {2019})}\BibitemShut {NoStop}%
\bibitem [{\citenamefont {Yokomizo}\ and\ \citenamefont
  {Murakami}(2019)}]{Yokomizo2019}%
  \BibitemOpen
  \bibfield  {author} {\bibinfo {author} {\bibfnamefont {K.}~\bibnamefont
  {Yokomizo}}\ and\ \bibinfo {author} {\bibfnamefont {S.}~\bibnamefont
  {Murakami}},\ }\href {\doibase 10.1103/PhysRevLett.123.066404} {\bibfield
  {journal} {\bibinfo  {journal} {Phys. Rev. Lett.}\ }\textbf {\bibinfo
  {volume} {123}},\ \bibinfo {pages} {066404} (\bibinfo {year}
  {2019})}\BibitemShut {NoStop}%
\bibitem [{\citenamefont {Okuma}\ and\ \citenamefont {Sato}(2019)}]{Okuma2019}%
  \BibitemOpen
  \bibfield  {author} {\bibinfo {author} {\bibfnamefont {N.}~\bibnamefont
  {Okuma}}\ and\ \bibinfo {author} {\bibfnamefont {M.}~\bibnamefont {Sato}},\
  }\href {\doibase 10.1103/PhysRevLett.123.097701} {\bibfield  {journal}
  {\bibinfo  {journal} {Phys. Rev. Lett.}\ }\textbf {\bibinfo {volume} {123}},\
  \bibinfo {pages} {097701} (\bibinfo {year} {2019})}\BibitemShut {NoStop}%
\bibitem [{\citenamefont {Wu}\ \emph {et~al.}(2019{\natexlab{b}})\citenamefont
  {Wu}, \citenamefont {Jin},\ and\ \citenamefont {Song}}]{HWu2019}%
  \BibitemOpen
  \bibfield  {author} {\bibinfo {author} {\bibfnamefont {H.~C.}\ \bibnamefont
  {Wu}}, \bibinfo {author} {\bibfnamefont {L.}~\bibnamefont {Jin}}, \ and\
  \bibinfo {author} {\bibfnamefont {Z.}~\bibnamefont {Song}},\ }\href {\doibase
  10.1103/PhysRevB.100.155117} {\bibfield  {journal} {\bibinfo  {journal}
  {Phys. Rev. B}\ }\textbf {\bibinfo {volume} {100}},\ \bibinfo {pages}
  {155117} (\bibinfo {year} {2019}{\natexlab{b}})}\BibitemShut {NoStop}%
\bibitem [{\citenamefont {Brzezicki}\ and\ \citenamefont
  {Hyart}(2019)}]{Brzezicki2019}%
  \BibitemOpen
  \bibfield  {author} {\bibinfo {author} {\bibfnamefont {W.}~\bibnamefont
  {Brzezicki}}\ and\ \bibinfo {author} {\bibfnamefont {T.}~\bibnamefont
  {Hyart}},\ }\href {\doibase 10.1103/PhysRevB.100.161105} {\bibfield
  {journal} {\bibinfo  {journal} {Phys. Rev. B}\ }\textbf {\bibinfo {volume}
  {100}},\ \bibinfo {pages} {161105(R)} (\bibinfo {year} {2019})}\BibitemShut
  {NoStop}%
\bibitem [{\citenamefont {Kawabata}\ \emph
  {et~al.}(2019{\natexlab{c}})\citenamefont {Kawabata}, \citenamefont
  {Shiozaki}, \citenamefont {Ueda},\ and\ \citenamefont
  {Sato}}]{Kawabata2019v3}%
  \BibitemOpen
  \bibfield  {author} {\bibinfo {author} {\bibfnamefont {K.}~\bibnamefont
  {Kawabata}}, \bibinfo {author} {\bibfnamefont {K.}~\bibnamefont {Shiozaki}},
  \bibinfo {author} {\bibfnamefont {M.}~\bibnamefont {Ueda}}, \ and\ \bibinfo
  {author} {\bibfnamefont {M.}~\bibnamefont {Sato}},\ }\href {\doibase
  10.1103/PhysRevX.9.041015} {\bibfield  {journal} {\bibinfo  {journal} {Phys.
  Rev. X}\ }\textbf {\bibinfo {volume} {9}},\ \bibinfo {pages} {041015}
  (\bibinfo {year} {2019}{\natexlab{c}})}\BibitemShut {NoStop}%
\bibitem [{\citenamefont {Borgnia}\ \emph {et~al.}(2020)\citenamefont
  {Borgnia}, \citenamefont {Kruchkov},\ and\ \citenamefont
  {Slager}}]{Borgnia2020}%
  \BibitemOpen
  \bibfield  {author} {\bibinfo {author} {\bibfnamefont {D.~S.}\ \bibnamefont
  {Borgnia}}, \bibinfo {author} {\bibfnamefont {A.~J.}\ \bibnamefont
  {Kruchkov}}, \ and\ \bibinfo {author} {\bibfnamefont {R.-J.}\ \bibnamefont
  {Slager}},\ }\href {\doibase 10.1103/PhysRevLett.124.056802} {\bibfield
  {journal} {\bibinfo  {journal} {Phys. Rev. Lett.}\ }\textbf {\bibinfo
  {volume} {124}},\ \bibinfo {pages} {056802} (\bibinfo {year}
  {2020})}\BibitemShut {NoStop}%
\bibitem [{\citenamefont {Okuma}\ \emph {et~al.}(2020)\citenamefont {Okuma},
  \citenamefont {Kawabata}, \citenamefont {Shiozaki},\ and\ \citenamefont
  {Sato}}]{Okuma2020}%
  \BibitemOpen
  \bibfield  {author} {\bibinfo {author} {\bibfnamefont {N.}~\bibnamefont
  {Okuma}}, \bibinfo {author} {\bibfnamefont {K.}~\bibnamefont {Kawabata}},
  \bibinfo {author} {\bibfnamefont {K.}~\bibnamefont {Shiozaki}}, \ and\
  \bibinfo {author} {\bibfnamefont {M.}~\bibnamefont {Sato}},\ }\href {\doibase
  10.1103/PhysRevLett.124.086801} {\bibfield  {journal} {\bibinfo  {journal}
  {Phys. Rev. Lett.}\ }\textbf {\bibinfo {volume} {124}},\ \bibinfo {pages}
  {086801} (\bibinfo {year} {2020})}\BibitemShut {NoStop}%
\bibitem [{\citenamefont {Yin}\ \emph {et~al.}(2018)\citenamefont {Yin},
  \citenamefont {Jiang}, \citenamefont {Li}, \citenamefont {L\"u},\ and\
  \citenamefont {Chen}}]{Yin2018}%
  \BibitemOpen
  \bibfield  {author} {\bibinfo {author} {\bibfnamefont {C.}~\bibnamefont
  {Yin}}, \bibinfo {author} {\bibfnamefont {H.}~\bibnamefont {Jiang}}, \bibinfo
  {author} {\bibfnamefont {L.}~\bibnamefont {Li}}, \bibinfo {author}
  {\bibfnamefont {R.}~\bibnamefont {L\"u}}, \ and\ \bibinfo {author}
  {\bibfnamefont {S.}~\bibnamefont {Chen}},\ }\href {\doibase
  10.1103/PhysRevA.97.052115} {\bibfield  {journal} {\bibinfo  {journal} {Phys.
  Rev. A}\ }\textbf {\bibinfo {volume} {97}},\ \bibinfo {pages} {052115}
  (\bibinfo {year} {2018})}\BibitemShut {NoStop}%
\bibitem [{\citenamefont {Herviou}\ \emph {et~al.}(2019)\citenamefont
  {Herviou}, \citenamefont {Bardarson},\ and\ \citenamefont
  {Regnault}}]{Herviou2019}%
  \BibitemOpen
  \bibfield  {author} {\bibinfo {author} {\bibfnamefont {L.}~\bibnamefont
  {Herviou}}, \bibinfo {author} {\bibfnamefont {J.~H.}\ \bibnamefont
  {Bardarson}}, \ and\ \bibinfo {author} {\bibfnamefont {N.}~\bibnamefont
  {Regnault}},\ }\href {\doibase 10.1103/PhysRevA.99.052118} {\bibfield
  {journal} {\bibinfo  {journal} {Phys. Rev. A}\ }\textbf {\bibinfo {volume}
  {99}},\ \bibinfo {pages} {052118} (\bibinfo {year} {2019})}\BibitemShut
  {NoStop}%
\bibitem [{\citenamefont {Wu}\ and\ \citenamefont {Hou}(2019)}]{YWu2019}%
  \BibitemOpen
  \bibfield  {author} {\bibinfo {author} {\bibfnamefont {Y.-J.}\ \bibnamefont
  {Wu}}\ and\ \bibinfo {author} {\bibfnamefont {J.}~\bibnamefont {Hou}},\
  }\href {\doibase 10.1103/PhysRevA.99.062107} {\bibfield  {journal} {\bibinfo
  {journal} {Phys. Rev. A}\ }\textbf {\bibinfo {volume} {99}},\ \bibinfo
  {pages} {062107} (\bibinfo {year} {2019})}\BibitemShut {NoStop}%
\bibitem [{\citenamefont {Chen}\ \emph
  {et~al.}(2019{\natexlab{b}})\citenamefont {Chen}, \citenamefont {Chen},
  \citenamefont {Zhou},\ and\ \citenamefont {Xu}}]{RChen2019}%
  \BibitemOpen
  \bibfield  {author} {\bibinfo {author} {\bibfnamefont {R.}~\bibnamefont
  {Chen}}, \bibinfo {author} {\bibfnamefont {C.-Z.}\ \bibnamefont {Chen}},
  \bibinfo {author} {\bibfnamefont {B.}~\bibnamefont {Zhou}}, \ and\ \bibinfo
  {author} {\bibfnamefont {D.-H.}\ \bibnamefont {Xu}},\ }\href {\doibase
  10.1103/PhysRevB.99.155431} {\bibfield  {journal} {\bibinfo  {journal} {Phys.
  Rev. B}\ }\textbf {\bibinfo {volume} {99}},\ \bibinfo {pages} {155431}
  (\bibinfo {year} {2019}{\natexlab{b}})}\BibitemShut {NoStop}%
\bibitem [{\citenamefont {Li}\ \emph {et~al.}(2019{\natexlab{b}})\citenamefont
  {Li}, \citenamefont {Lee},\ and\ \citenamefont {Gong}}]{LLi2019}%
  \BibitemOpen
  \bibfield  {author} {\bibinfo {author} {\bibfnamefont {L.}~\bibnamefont
  {Li}}, \bibinfo {author} {\bibfnamefont {C.~H.}\ \bibnamefont {Lee}}, \ and\
  \bibinfo {author} {\bibfnamefont {J.}~\bibnamefont {Gong}},\ }\href {\doibase
  10.1103/PhysRevB.100.075403} {\bibfield  {journal} {\bibinfo  {journal}
  {Phys. Rev. B}\ }\textbf {\bibinfo {volume} {100}},\ \bibinfo {pages}
  {075403} (\bibinfo {year} {2019}{\natexlab{b}})}\BibitemShut {NoStop}%
\bibitem [{\citenamefont {Loic}\ \emph {et~al.}(2019)\citenamefont {Loic},
  \citenamefont {Nicolas},\ and\ \citenamefont {Jens}}]{Loic2019}%
  \BibitemOpen
  \bibfield  {author} {\bibinfo {author} {\bibfnamefont {H.}~\bibnamefont
  {Loic}}, \bibinfo {author} {\bibfnamefont {R.}~\bibnamefont {Nicolas}}, \
  and\ \bibinfo {author} {\bibfnamefont {H.~B.}\ \bibnamefont {Jens}},\ }\href
  {\doibase 10.21468/SciPostPhys.7.5.069} {\bibfield  {journal} {\bibinfo
  {journal} {SciPost Phys.}\ }\textbf {\bibinfo {volume} {7}},\ \bibinfo
  {pages} {69} (\bibinfo {year} {2019})}\BibitemShut {NoStop}%
\bibitem [{\citenamefont {Wang}\ \emph {et~al.}(2020)\citenamefont {Wang},
  \citenamefont {Guo},\ and\ \citenamefont {Kou}}]{Wang2020}%
  \BibitemOpen
  \bibfield  {author} {\bibinfo {author} {\bibfnamefont {X.-R.}\ \bibnamefont
  {Wang}}, \bibinfo {author} {\bibfnamefont {C.-X.}\ \bibnamefont {Guo}}, \
  and\ \bibinfo {author} {\bibfnamefont {S.-P.}\ \bibnamefont {Kou}},\ }\href
  {\doibase 10.1103/PhysRevB.101.121116} {\bibfield  {journal} {\bibinfo
  {journal} {Phys. Rev. B}\ }\textbf {\bibinfo {volume} {101}},\ \bibinfo
  {pages} {121116} (\bibinfo {year} {2020})}\BibitemShut {NoStop}%
\bibitem [{\citenamefont {Rudner}\ and\ \citenamefont
  {Levitov}(2009)}]{Runder2009}%
  \BibitemOpen
  \bibfield  {author} {\bibinfo {author} {\bibfnamefont {M.~S.}\ \bibnamefont
  {Rudner}}\ and\ \bibinfo {author} {\bibfnamefont {L.~S.}\ \bibnamefont
  {Levitov}},\ }\href {\doibase 10.1103/PhysRevLett.102.065703} {\bibfield
  {journal} {\bibinfo  {journal} {Phys. Rev. Lett.}\ }\textbf {\bibinfo
  {volume} {102}},\ \bibinfo {pages} {065703} (\bibinfo {year}
  {2009})}\BibitemShut {NoStop}%
\bibitem [{\citenamefont {Liang}\ and\ \citenamefont
  {Huang}(2013)}]{Liang2013}%
  \BibitemOpen
  \bibfield  {author} {\bibinfo {author} {\bibfnamefont {S.-D.}\ \bibnamefont
  {Liang}}\ and\ \bibinfo {author} {\bibfnamefont {G.-Y.}\ \bibnamefont
  {Huang}},\ }\href {\doibase 10.1103/PhysRevA.87.012118} {\bibfield  {journal}
  {\bibinfo  {journal} {Phys. Rev. A}\ }\textbf {\bibinfo {volume} {87}},\
  \bibinfo {pages} {012118} (\bibinfo {year} {2013})}\BibitemShut {NoStop}%
\bibitem [{\citenamefont {Zhu}\ \emph {et~al.}(2014)\citenamefont {Zhu},
  \citenamefont {L\"u},\ and\ \citenamefont {Chen}}]{Zhu2014}%
  \BibitemOpen
  \bibfield  {author} {\bibinfo {author} {\bibfnamefont {B.}~\bibnamefont
  {Zhu}}, \bibinfo {author} {\bibfnamefont {R.}~\bibnamefont {L\"u}}, \ and\
  \bibinfo {author} {\bibfnamefont {S.}~\bibnamefont {Chen}},\ }\href {\doibase
  10.1103/PhysRevA.89.062102} {\bibfield  {journal} {\bibinfo  {journal} {Phys.
  Rev. A}\ }\textbf {\bibinfo {volume} {89}},\ \bibinfo {pages} {062102}
  (\bibinfo {year} {2014})}\BibitemShut {NoStop}%
\bibitem [{\citenamefont {Zhao}\ \emph {et~al.}(2015)\citenamefont {Zhao},
  \citenamefont {Longhi},\ and\ \citenamefont {Feng}}]{Zhao2015}%
  \BibitemOpen
  \bibfield  {author} {\bibinfo {author} {\bibfnamefont {H.}~\bibnamefont
  {Zhao}}, \bibinfo {author} {\bibfnamefont {S.}~\bibnamefont {Longhi}}, \ and\
  \bibinfo {author} {\bibfnamefont {L.}~\bibnamefont {Feng}},\ }\href@noop {}
  {\bibfield  {journal} {\bibinfo  {journal} {Sci. Rep.}\ }\textbf {\bibinfo
  {volume} {5}},\ \bibinfo {pages} {17022} (\bibinfo {year}
  {2015})}\BibitemShut {NoStop}%
\bibitem [{\citenamefont {Jin}\ \emph {et~al.}(2017)\citenamefont {Jin},
  \citenamefont {Wang},\ and\ \citenamefont {Song}}]{Jin2017}%
  \BibitemOpen
  \bibfield  {author} {\bibinfo {author} {\bibfnamefont {L.}~\bibnamefont
  {Jin}}, \bibinfo {author} {\bibfnamefont {P.}~\bibnamefont {Wang}}, \ and\
  \bibinfo {author} {\bibfnamefont {Z.}~\bibnamefont {Song}},\ }\href@noop {}
  {\bibfield  {journal} {\bibinfo  {journal} {Sci. Rep.}\ }\textbf {\bibinfo
  {volume} {7}},\ \bibinfo {pages} {5903} (\bibinfo {year} {2017})}\BibitemShut
  {NoStop}%
\bibitem [{\citenamefont {Yuce}(2018)}]{Yuce2018pra}%
  \BibitemOpen
  \bibfield  {author} {\bibinfo {author} {\bibfnamefont {C.}~\bibnamefont
  {Yuce}},\ }\href {\doibase 10.1103/PhysRevA.97.042118} {\bibfield  {journal}
  {\bibinfo  {journal} {Phys. Rev. A}\ }\textbf {\bibinfo {volume} {97}},\
  \bibinfo {pages} {042118} (\bibinfo {year} {2018})}\BibitemShut {NoStop}%
\bibitem [{\citenamefont {Lieu}(2018)}]{Lieu2018}%
  \BibitemOpen
  \bibfield  {author} {\bibinfo {author} {\bibfnamefont {S.}~\bibnamefont
  {Lieu}},\ }\href {\doibase 10.1103/PhysRevB.97.045106} {\bibfield  {journal}
  {\bibinfo  {journal} {Phys. Rev. B}\ }\textbf {\bibinfo {volume} {97}},\
  \bibinfo {pages} {045106} (\bibinfo {year} {2018})}\BibitemShut {NoStop}%
\bibitem [{\citenamefont {Klett}\ \emph {et~al.}(2018)\citenamefont {Klett},
  \citenamefont {Cartarius}, \citenamefont {Dast}, \citenamefont {Main},\ and\
  \citenamefont {Wunner}}]{Klett2018}%
  \BibitemOpen
  \bibfield  {author} {\bibinfo {author} {\bibfnamefont {M.}~\bibnamefont
  {Klett}}, \bibinfo {author} {\bibfnamefont {H.}~\bibnamefont {Cartarius}},
  \bibinfo {author} {\bibfnamefont {D.}~\bibnamefont {Dast}}, \bibinfo {author}
  {\bibfnamefont {J.}~\bibnamefont {Main}}, \ and\ \bibinfo {author}
  {\bibfnamefont {G.}~\bibnamefont {Wunner}},\ }\href@noop {} {\bibfield
  {journal} {\bibinfo  {journal} {Eur. Phys. J. D}\ }\textbf {\bibinfo {volume}
  {72}},\ \bibinfo {pages} {214} (\bibinfo {year} {2018})}\BibitemShut
  {NoStop}%
\bibitem [{\citenamefont {Yuce}\ and\ \citenamefont {Oztas}(2018)}]{Yuce2018}%
  \BibitemOpen
  \bibfield  {author} {\bibinfo {author} {\bibfnamefont {C.}~\bibnamefont
  {Yuce}}\ and\ \bibinfo {author} {\bibfnamefont {Z.}~\bibnamefont {Oztas}},\
  }\href@noop {} {\bibfield  {journal} {\bibinfo  {journal} {Sci. Rep.}\
  }\textbf {\bibinfo {volume} {8}},\ \bibinfo {pages} {17416} (\bibinfo {year}
  {2018})}\BibitemShut {NoStop}%
\bibitem [{\citenamefont {Ge}\ \emph {et~al.}(2019)\citenamefont {Ge},
  \citenamefont {Zhang}, \citenamefont {Liu}, \citenamefont {Li}, \citenamefont
  {Fan},\ and\ \citenamefont {Nori}}]{Ge2019}%
  \BibitemOpen
  \bibfield  {author} {\bibinfo {author} {\bibfnamefont {Z.-Y.}\ \bibnamefont
  {Ge}}, \bibinfo {author} {\bibfnamefont {Y.-R.}\ \bibnamefont {Zhang}},
  \bibinfo {author} {\bibfnamefont {T.}~\bibnamefont {Liu}}, \bibinfo {author}
  {\bibfnamefont {S.-W.}\ \bibnamefont {Li}}, \bibinfo {author} {\bibfnamefont
  {H.}~\bibnamefont {Fan}}, \ and\ \bibinfo {author} {\bibfnamefont
  {F.}~\bibnamefont {Nori}},\ }\href {\doibase 10.1103/PhysRevB.100.054105}
  {\bibfield  {journal} {\bibinfo  {journal} {Phys. Rev. B}\ }\textbf {\bibinfo
  {volume} {100}},\ \bibinfo {pages} {054105} (\bibinfo {year}
  {2019})}\BibitemShut {NoStop}%
\bibitem [{\citenamefont {Fu}\ \emph {et~al.}(2020)\citenamefont {Fu},
  \citenamefont {Fu}, \citenamefont {Zhang}, \citenamefont {Wang},
  \citenamefont {Zhao},\ and\ \citenamefont {Ke}}]{Fu2020}%
  \BibitemOpen
  \bibfield  {author} {\bibinfo {author} {\bibfnamefont {Z.}~\bibnamefont
  {Fu}}, \bibinfo {author} {\bibfnamefont {N.}~\bibnamefont {Fu}}, \bibinfo
  {author} {\bibfnamefont {H.}~\bibnamefont {Zhang}}, \bibinfo {author}
  {\bibfnamefont {Z.}~\bibnamefont {Wang}}, \bibinfo {author} {\bibfnamefont
  {D.}~\bibnamefont {Zhao}}, \ and\ \bibinfo {author} {\bibfnamefont
  {S.}~\bibnamefont {Ke}},\ }\href@noop {} {\bibfield  {journal} {\bibinfo
  {journal} {Appl. Sci.}\ }\textbf {\bibinfo {volume} {10}},\ \bibinfo {pages}
  {3425} (\bibinfo {year} {2020})}\BibitemShut {NoStop}%
\bibitem [{\citenamefont {Kunst}\ \emph {et~al.}(2018)\citenamefont {Kunst},
  \citenamefont {Edvardsson}, \citenamefont {Budich},\ and\ \citenamefont
  {Bergholtz}}]{Kunst2018}%
  \BibitemOpen
  \bibfield  {author} {\bibinfo {author} {\bibfnamefont {F.~K.}\ \bibnamefont
  {Kunst}}, \bibinfo {author} {\bibfnamefont {E.}~\bibnamefont {Edvardsson}},
  \bibinfo {author} {\bibfnamefont {J.~C.}\ \bibnamefont {Budich}}, \ and\
  \bibinfo {author} {\bibfnamefont {E.~J.}\ \bibnamefont {Bergholtz}},\ }\href
  {\doibase 10.1103/PhysRevLett.121.026808} {\bibfield  {journal} {\bibinfo
  {journal} {Phys. Rev. Lett.}\ }\textbf {\bibinfo {volume} {121}},\ \bibinfo
  {pages} {026808} (\bibinfo {year} {2018})}\BibitemShut {NoStop}%
\bibitem [{\citenamefont {Yao}\ and\ \citenamefont {Wang}(2018)}]{Yao2018}%
  \BibitemOpen
  \bibfield  {author} {\bibinfo {author} {\bibfnamefont {S.}~\bibnamefont
  {Yao}}\ and\ \bibinfo {author} {\bibfnamefont {Z.}~\bibnamefont {Wang}},\
  }\href {\doibase 10.1103/PhysRevLett.121.086803} {\bibfield  {journal}
  {\bibinfo  {journal} {Phys. Rev. Lett.}\ }\textbf {\bibinfo {volume} {121}},\
  \bibinfo {pages} {086803} (\bibinfo {year} {2018})}\BibitemShut {NoStop}%
\bibitem [{\citenamefont {Jin}\ and\ \citenamefont {Song}(2019)}]{Jin2019}%
  \BibitemOpen
  \bibfield  {author} {\bibinfo {author} {\bibfnamefont {L.}~\bibnamefont
  {Jin}}\ and\ \bibinfo {author} {\bibfnamefont {Z.}~\bibnamefont {Song}},\
  }\href {\doibase 10.1103/PhysRevB.99.081103} {\bibfield  {journal} {\bibinfo
  {journal} {Phys. Rev. B}\ }\textbf {\bibinfo {volume} {99}},\ \bibinfo
  {pages} {081103(R)} (\bibinfo {year} {2019})}\BibitemShut {NoStop}%
\bibitem [{\citenamefont {Lee}\ and\ \citenamefont {Thomale}(2019)}]{Lee2019}%
  \BibitemOpen
  \bibfield  {author} {\bibinfo {author} {\bibfnamefont {C.~H.}\ \bibnamefont
  {Lee}}\ and\ \bibinfo {author} {\bibfnamefont {R.}~\bibnamefont {Thomale}},\
  }\href {\doibase 10.1103/PhysRevB.99.201103} {\bibfield  {journal} {\bibinfo
  {journal} {Phys. Rev. B}\ }\textbf {\bibinfo {volume} {99}},\ \bibinfo
  {pages} {201103(R)} (\bibinfo {year} {2019})}\BibitemShut {NoStop}%
\bibitem [{\citenamefont {Kunst}\ and\ \citenamefont
  {Dwivedi}(2019)}]{Kunst2019}%
  \BibitemOpen
  \bibfield  {author} {\bibinfo {author} {\bibfnamefont {F.~K.}\ \bibnamefont
  {Kunst}}\ and\ \bibinfo {author} {\bibfnamefont {V.}~\bibnamefont
  {Dwivedi}},\ }\href {\doibase 10.1103/PhysRevB.99.245116} {\bibfield
  {journal} {\bibinfo  {journal} {Phys. Rev. B}\ }\textbf {\bibinfo {volume}
  {99}},\ \bibinfo {pages} {245116} (\bibinfo {year} {2019})}\BibitemShut
  {NoStop}%
\bibitem [{\citenamefont {Deng}\ and\ \citenamefont {Yi}(2019)}]{Deng2019}%
  \BibitemOpen
  \bibfield  {author} {\bibinfo {author} {\bibfnamefont {T.-S.}\ \bibnamefont
  {Deng}}\ and\ \bibinfo {author} {\bibfnamefont {W.}~\bibnamefont {Yi}},\
  }\href {\doibase 10.1103/PhysRevB.100.035102} {\bibfield  {journal} {\bibinfo
   {journal} {Phys. Rev. B}\ }\textbf {\bibinfo {volume} {100}},\ \bibinfo
  {pages} {035102} (\bibinfo {year} {2019})}\BibitemShut {NoStop}%
\bibitem [{\citenamefont {Imura}\ and\ \citenamefont
  {Takane}(2019)}]{Imura2019}%
  \BibitemOpen
  \bibfield  {author} {\bibinfo {author} {\bibfnamefont {K.-I.}\ \bibnamefont
  {Imura}}\ and\ \bibinfo {author} {\bibfnamefont {Y.}~\bibnamefont {Takane}},\
  }\href {\doibase 10.1103/PhysRevB.100.165430} {\bibfield  {journal} {\bibinfo
   {journal} {Phys. Rev. B}\ }\textbf {\bibinfo {volume} {100}},\ \bibinfo
  {pages} {165430} (\bibinfo {year} {2019})}\BibitemShut {NoStop}%
\bibitem [{\citenamefont {Song}\ \emph
  {et~al.}(2019{\natexlab{a}})\citenamefont {Song}, \citenamefont {Yao},\ and\
  \citenamefont {Wang}}]{Song2019}%
  \BibitemOpen
  \bibfield  {author} {\bibinfo {author} {\bibfnamefont {F.}~\bibnamefont
  {Song}}, \bibinfo {author} {\bibfnamefont {S.}~\bibnamefont {Yao}}, \ and\
  \bibinfo {author} {\bibfnamefont {Z.}~\bibnamefont {Wang}},\ }\href {\doibase
  10.1103/PhysRevLett.123.170401} {\bibfield  {journal} {\bibinfo  {journal}
  {Phys. Rev. Lett.}\ }\textbf {\bibinfo {volume} {123}},\ \bibinfo {pages}
  {170401} (\bibinfo {year} {2019}{\natexlab{a}})}\BibitemShut {NoStop}%
\bibitem [{\citenamefont {Song}\ \emph
  {et~al.}(2019{\natexlab{b}})\citenamefont {Song}, \citenamefont {Yao},\ and\
  \citenamefont {Wang}}]{Song2019v2}%
  \BibitemOpen
  \bibfield  {author} {\bibinfo {author} {\bibfnamefont {F.}~\bibnamefont
  {Song}}, \bibinfo {author} {\bibfnamefont {S.}~\bibnamefont {Yao}}, \ and\
  \bibinfo {author} {\bibfnamefont {Z.}~\bibnamefont {Wang}},\ }\href {\doibase
  10.1103/PhysRevLett.123.246801} {\bibfield  {journal} {\bibinfo  {journal}
  {Phys. Rev. Lett.}\ }\textbf {\bibinfo {volume} {123}},\ \bibinfo {pages}
  {246801} (\bibinfo {year} {2019}{\natexlab{b}})}\BibitemShut {NoStop}%
\bibitem [{\citenamefont {Longhi}(2019)}]{Longhi2019}%
  \BibitemOpen
  \bibfield  {author} {\bibinfo {author} {\bibfnamefont {S.}~\bibnamefont
  {Longhi}},\ }\href {\doibase 10.1103/PhysRevResearch.1.023013} {\bibfield
  {journal} {\bibinfo  {journal} {Phys. Rev. Research}\ }\textbf {\bibinfo
  {volume} {1}},\ \bibinfo {pages} {023013} (\bibinfo {year}
  {2019})}\BibitemShut {NoStop}%
\bibitem [{\citenamefont {Zhang}\ \emph {et~al.}(2020)\citenamefont {Zhang},
  \citenamefont {Yang},\ and\ \citenamefont {Fang}}]{KZhang2020}%
  \BibitemOpen
  \bibfield  {author} {\bibinfo {author} {\bibfnamefont {K.}~\bibnamefont
  {Zhang}}, \bibinfo {author} {\bibfnamefont {Z.}~\bibnamefont {Yang}}, \ and\
  \bibinfo {author} {\bibfnamefont {C.}~\bibnamefont {Fang}},\ }\href {\doibase
  10.1103/PhysRevLett.125.126402} {\bibfield  {journal} {\bibinfo  {journal}
  {Phys. Rev. Lett.}\ }\textbf {\bibinfo {volume} {125}},\ \bibinfo {pages}
  {126402} (\bibinfo {year} {2020})}\BibitemShut {NoStop}%
\bibitem [{\citenamefont {Yang}\ \emph {et~al.}()\citenamefont {Yang},
  \citenamefont {Zhang}, \citenamefont {Fang},\ and\ \citenamefont
  {Hu}}]{ZYang2019}%
  \BibitemOpen
  \bibfield  {author} {\bibinfo {author} {\bibfnamefont {Z.}~\bibnamefont
  {Yang}}, \bibinfo {author} {\bibfnamefont {K.}~\bibnamefont {Zhang}},
  \bibinfo {author} {\bibfnamefont {C.}~\bibnamefont {Fang}}, \ and\ \bibinfo
  {author} {\bibfnamefont {J.}~\bibnamefont {Hu}},\ }\href@noop {} {\bibinfo
  {journal} {arXiv:1912.05499}\ }\BibitemShut {NoStop}%
\bibitem [{\citenamefont {Kitaev}(2001)}]{Kitaev2001}%
  \BibitemOpen
\bibfield  {journal} {  }\bibfield  {author} {\bibinfo {author} {\bibfnamefont
  {A.~Y.}\ \bibnamefont {Kitaev}},\ }\href {\doibase
  10.1070/1063-7869/44/10s/s29} {\bibfield  {journal} {\bibinfo  {journal}
  {Phys. Usp.}\ }\textbf {\bibinfo {volume} {44}},\ \bibinfo {pages} {131}
  (\bibinfo {year} {2001})}\BibitemShut {NoStop}%
\end{thebibliography}
\end{document}